%% file: main.tex
\documentclass[letterpaper,journal]{IEEEtran}

\input{preamble}

\begin{document}

% ---------- 标题 ----------
\title{WAFT: A Fourier-Analytic Framework for Steady-State Performance of Walker Constellations}

% ---------- 作者与单位 ----------
\author{Haoyang~Xue,~\IEEEmembership{Student Member,~IEEE,}
        Bingkun~Liu,~\IEEEmembership{Member,~IEEE,}
        and~Linling~Kuang,~\IEEEmembership{Senior Member,~IEEE}
\thanks{This work was supported by the Beijing Natural Science Foundation (L244007) and the Project Financially Supported by Laoshan Laboratory (No. LSKJ202600900). \textit{(Corresponding author: Linling Kuang.)}}%
\thanks{Haoyang Xue is with the Department of Electronic Engineering, Tsinghua University, Beijing, China (e-mail: xuehy24@mails.tsinghua.edu.cn).}%
\thanks{Bingkun Liu is with the Beijing National Research Center for Information Science and Technology, Tsinghua University, Beijing 100084, China (e-mail: lbk@mail.tsinghua.edu.cn).}%
\thanks{Linling Kuang is with the State Key Laboratory of Space Network and Communications, Tsinghua University, Beijing 100084, China (e-mail: kll@tsinghua.edu.cn).}}

% ---------- 页眉 ----------
% 第一参数：稿件状态（投稿版用 "Submitted to ..."，不要写 Preprint）；第二参数：作者 + 标题
\markboth{Submitted to IEEE Transactions on Wireless Communications}%
{Xue \MakeLowercase{\textit{et al.}}: WAFT: A Fourier-Analytic Framework for Steady-State Performance of Walker Constellations}

\maketitle

% ---------- arXiv 预印本声明（IEEE PSPB 8.1.10B 要求，仅 arXiv 版保留） ----------
% \enlargethispage 让首页多容纳与声明等高的正文，避免声明把后续浮动体整体推迟一页
\enlargethispage{2\baselineskip}
\begin{center}
\footnotesize This work has been submitted to the IEEE for possible publication.\\
Copyright may be transferred without notice, after which this version may no longer be accessible.
\end{center}

% ---------- 摘要 + 关键词 ----------
\input{sections/abstract}

% ---------- 正文各节 ----------
\input{sections/intro}
\input{sections/system_model}

\input{sections/performance_analysis}

\input{sections/interference}

% 在结论前设置浮动图边界，防止结果图漂入附录。
\FloatBarrier
\input{sections/conclusion}

% ---------- 附录 ----------
% IEEEtran 用 \appendices（复数）：其后每个 \section 自动编号为 APPENDIX A/B/…；
% 附录内的公式默认「延续正文编号」（如需 (A1) 形式需另行设置）。
\appendices
\input{sections/appendix}

% ---------- 参考文献 ----------
% \balance 使末页两栏高度平衡（来自 balance 宏包）
\balance
\bibliographystyle{IEEEtran}
\bibliography{refs}

\end{document}

%% file: preamble.tex
\usepackage{amsmath,amsfonts}
\usepackage{mathtools}       % multlined：多行式子内部块（首行左齐、末行右齐）
\usepackage{amssymb}         % \triangleq 等 AMS 符号
\usepackage{dsfont}          % \mathds{1}：示性函数（黑板粗体 1）
\usepackage{algorithm}
\usepackage{algorithmic}
\usepackage{array}
\usepackage[caption=false]{subfig}
\usepackage{textcomp}
\usepackage{stfloats}
\usepackage{float}
\usepackage{placeins}
\usepackage{url}
\usepackage{verbatim}
\usepackage{graphicx}
\usepackage{cite}
\usepackage{balance}

\def\IEEEbibitemsep{0pt}

\newtheorem{lemma}{Lemma}
\newtheorem{theorem}{Theorem}
\newtheorem{corollary}{Corollary}
\newtheorem{definition}{Definition}
\newtheorem{remark}{Remark}

\graphicspath{{figures/}}

\usepackage[hidelinks,bookmarksnumbered]{hyperref}
\hypersetup{
    pdfcreator={LaTeX with IEEEtran},
    pdfproducer={pdfTeX},
}

%% file: sections/abstract.tex
\begin{abstract}
The Walker configuration is predominantly adopted by nongeostationary orbit constellations exemplified by Starlink and Kuiper.
The steady-state distribution of aggregate downlink interference power is critical for spectrum coexistence regulation.
Large-scale time-stepped simulation and stochastic geometry (SG)-based models are the mainstream methods for obtaining the distribution, but are either computationally expensive or subject to inherent bias.
This paper models the aggregate downlink interference power as an additive observable and proposes the Walker-constellation Analysis via Fourier Transform (WAFT) framework to derive the steady-state performance.
Modeling the constellation as a periodic impulse array on the phase torus, the spectral representation of the observable is derived via 2-D Fourier analysis.
Analytical expressions for the expectation, variance, and steady-state complementary cumulative distribution function are further derived, and the periodicity of receiver performance along longitude is characterized.
For a Starlink sub-constellation example, numerical validation shows that the percentile discrepancies of WAFT relative to a finite-duration time-stepped simulation remain within 0.09~dB.
Comparisons with isotropic and intensity-matched SG models reveal that the conditional independence of satellite positions alone makes the models assign a substantial probability mass to physically infeasible values, producing percentile discrepancies.
\end{abstract}


%% file: sections/intro.tex
\section{Introduction}
\label{sec:intro}

\IEEEPARstart{T}{he} Walker configuration is the predominant architecture of nongeostationary orbit (NGSO) constellations, exemplified by Starlink and Kuiper. Such constellations are reshaping global broadband access and have become critical infrastructure for sixth-generation (6G) integrated space-terrestrial networks \cite{Arch_KeyTech,NewSpaceEra,FiveGFrmSpce}. 
A Walker-$\delta$ constellation \cite{walker1970_circular} is composed of $N_o$ orbital planes equally spaced in longitude of the ascending node (LAN) and $N_s$ satellites equally spaced within each plane, with the relative phasing between adjacent planes locked by a phasing factor $F$.

On the regulatory front, the focus is shifting from the expectation of aggregate downlink interference power toward percentiles of long-term statistics, i.e., equivalent power flux-density (EPFD) thresholds read off the steady-state complementary cumulative distribution function (CCDF) to protect geostationary orbit (GSO) system receivers \cite{itu_rr2024_art22, itu2024_wrc23}.
For a terrestrial receiver, however, the field of view (FOV) is a rotationally symmetric spherical cap containing a highly nonuniform and continuously time-varying subset of the constellation, rendering steady-state performance difficult to characterize through direct geometric analysis alone.

Consequently, large-scale time-stepped simulation is commonly used to obtain the steady-state distribution.
Constellation orbits are propagated to evaluate performance at each time instant, and the distribution is approximated from the resulting finite-duration statistics \cite{CmparOf3LEO,NewCmparOf4LEO,DelyNotOption}. 
The ITU-R S.1503 algorithm is used to verify EPFD compliance for spectrum regulation \cite{itur2023s1503}. 
Such simulation closely matches real systems but does not directly reveal how performance depends on constellation configuration, and the computational cost grows rapidly with constellation size and the number of parameter combinations.
Distribution-oriented alternatives likewise return prescribed metrics \cite{Kuang_AggInt_AAST, Kuang_UplinkInt_IoTJ, Kuang_BlindSpot_TAES}.

Stochastic geometry (SG) is a probabilistic framework \cite{SG_4WN} for approximating the constellation as a stochastic process. 
By modeling a constellation snapshot as a point process in which satellite positions are conditionally independent given a prescribed spatial structure \cite{SG_Survey}, steady-state performance metrics such as the nearest-satellite distance distribution and the aggregate downlink interference distribution can be analytically derived, and the dependence of performance on constellation parameters can be revealed. 
The earliest models adopt the spherical Poisson point process (PPP) or the binomial point process (BPP) \cite{SG_OtakiSG,SG_alhourani2021analytic,SG_BPPNearest,SG_4LEO,SG_Shadow,SG_wang2022conditional}, assuming uniformly distributed satellite positions to obtain simple and tractable results. 
Such models, however, do not explicitly represent the spatial correlation and nonuniformity of a regular constellation, thereby producing non-negligible bias. 
Later studies have sought to incorporate the structural features through a spatially varying intensity, i.e., the nonhomogeneous Poisson point process (NHPP) \cite{SG_OtakiNHPP,SG_Otaki_model_NHPP}, nested orbit-plane and in-plane point processes known as Cox point process (CPP) \cite{SG_Choi_HetLEO,SG_CB_CPP,SG_CB_CPP4HetLEO}, realistic orbital parameters \cite{SG_OrbGeo}, or inter-satellite spacing constraints \cite{SG_constraint_angle,SG_PAHPP}.
Nevertheless, the bias induced by the conditional independence of satellite positions is not fully removed \cite{SG_evaluate}, and the structural refinements that reduce it generally involve a trade-off between model realism and analytical tractability.
Moreover, being formulated in terms of spatial snapshots, SG-based models do not explicitly represent the temporal variation of the constellation or the temporal correlation of performance.

The Walker point process (WPP) \cite{CB_WPP_1} models a Walker constellation as a deterministic phase lattice with a random global phase offset, without assuming independence of satellite positions.
Exploiting the linear-flow dynamics on the phase torus spanned by the LAN and the in-plane phase, exact integral expressions are derived for key metrics such as the nearest-satellite distance, interference, and coverage probability.
The model is further extended to a multi-configuration constellation of superposed heterogeneous WPPs \cite{CB_WPP_2}.
The WPP models can capture the spatial structure of a Walker constellation accurately.
Existing WPP formulations, however, focus primarily on nonrepeating ground-track (non-RGT) constellations with $F=0$ \cite{CB_WPP_1,CB_WPP_2}. The influence of the phasing factor on performance and the spatiotemporal periodicity unique to repeating ground-track (RGT) constellations remain to be explored, and a unified analysis framework for general Walker constellations is not yet available.

This paper studies the steady-state performance of the aggregate interference experienced by a terrestrial receiver at an arbitrary location, for a single-layer Walker-$\delta$ constellation with circular orbits over an ideal spherical Earth.
Nonideal effects such as solar radiation pressure are neglected.
The receiver antenna is assumed static, with a time-invariant gain pattern and pointing.
The main contributions are summarized as follows:

\begin{itemize}
    \item The downlink aggregate interference power is modeled as an additive observable, and a unified framework termed Walker-constellation analysis via Fourier transform (WAFT) is developed for the steady-state analysis of additive observables in general Walker constellations.
    WAFT reduces the steady-state analysis of dynamically evolving Walker constellations to fixed spectral sampling within an invariant phase-space framework, enabling direct computation of steady-state statistics with explicitly controllable spectral truncation errors.
    \item The spectral representation of the observables is derived. This representation decouples the constellation structure, observable form, and temporal evolution.
    Changing Walker pattern only relocates the spectral samples, changing the observable form only modifies the underlying spatial spectrum, and advancing time only rotates the phases of constant-modulus complex exponentials, thereby enabling independent parametric analysis and unified performance evaluation across observables, constellation configurations, and time.
    \item Under explicit nondegeneracy conditions, closed-form minimum positive longitude periods are derived for the instantaneous performance in general Walker constellations and the steady-state performance in RGT constellations, the former proved an integer multiple of the latter. For RGT constellations, a nontrivial temporal period of the observables is derived, an integer fraction of the ground-track repeat period. These exact symmetries confine evaluation to fundamental temporal and longitude intervals and underpin longitude-dependent interference assessment and coordination.
    \item The WAFT framework is applied to downlink aggregate interference power and validated against finite-duration time-stepped simulations, with BPP, CPP, and NHPP as baselines. For the non-RGT Starlink example, the WAFT percentile discrepancies relative to the simulation are within 0.09~dB. The baseline comparison identifies the conditional independence of satellite positions as an important source of the SG discrepancy that persists under intensity matching.
\end{itemize}

The remainder of this paper is organized as follows.
The system model is established and additive observable is defined in Section~\ref{sec:system}.
The WAFT framework is developed and steady-state performance of additive observables is analyzed in Section~\ref{sec:performance}.
The framework is applied to the aggregate downlink interference in Section~\ref{sec:interference}, and the analytical results are validated against time-stepped simulations for both non-RGT and RGT constellations, with the BPP, CPP, and NHPP models as SG-based benchmarks.
The paper is concluded in Section~\ref{sec:conclusion} with future directions.

%% file: sections/system_model.tex
\section{System Model}
\label{sec:system}

\begin{figure}[!t]
    \centering
    \includegraphics{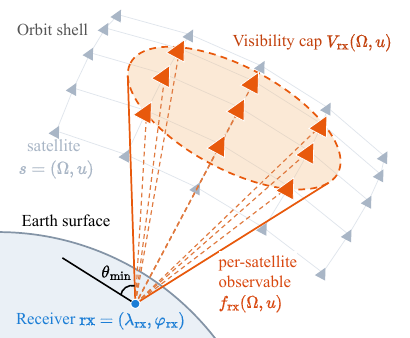}
    \caption{Illustration of the considered scenario.}
    \label{fig:scene}
\end{figure}

Fig.~\ref{fig:scene} illustrates the scenario. 
A terrestrial receiver observes the constellation through a view cone limited by a minimum elevation angle, which selects a visibility cap on the orbit shell such that only satellites inside the cap contribute to the performance.

\subsection{Walker Constellations and Dynamics in the Phase Torus}
\label{sec:walker-torus}

\subsubsection{Walker-\texorpdfstring{$\delta$}{delta} Constellations and the Phase Torus}

Consider an ideal single-layer circular-orbit Walker-$\delta$ constellation, commonly denoted by $i\!:\!N/N_o/F$ \cite{walker1970_circular}, where $i$ is the orbital inclination, $N$ the total number of satellites, $N_o$ the number of orbital planes, and $F\in\{0,\dots,N_o-1\}$ the phasing factor.
The $N_o$ planes are equally spaced in the longitude of the ascending node (LAN), each populated with $N_s=N/N_o$ equally spaced satellites.
All satellites share the same altitude $h$, giving an orbital radius $r = R_e + h$, where $R_e$ denotes the Earth's radius.

An Earth-Centered Earth-Fixed (ECEF) coordinate system is adopted.
Let $\Omega$ and $u$ denote the LAN and the in-plane phase, respectively, so that $(\Omega,u)$ constitutes the 2-D phase coordinate of a satellite.
The 3-D unit position vector of a satellite with phase coordinate $(\Omega, u)$ is
\begin{equation}
    \label{coordinates}
    \hat{\boldsymbol{r}}_{\mathrm{sat}}(\Omega, u)
    =
    \begin{pmatrix}
    \cos\Omega \cos u - \sin\Omega \cos i \sin u \\
    \sin\Omega \cos u + \cos\Omega \cos i \sin u \\
    \sin i \sin u
    \end{pmatrix}.
\end{equation}

The 3-D position of any satellite is a function of the 2-D phase coordinate, and the phase flow reproduces the orbital motion exactly, so constellation properties and their evolution can be studied in the phase domain without loss of information.
Since both $\Omega$ and $u$ are $2\pi$-periodic, $(\Omega,u)$ takes values on the 2-D torus $\mathbb{T}^2 \cong [0, 2\pi) \times [0, 2\pi)$, referred to as the \textit{phase torus}.

Without loss of generality, the zeroth satellite in the zeroth orbital plane is designated as the reference satellite, whose phase coordinate $\boldsymbol{x} = (\Omega_x, u_x)\in\mathbb{T}^2$ is referred to as the \textit{reference phase} of the constellation.

By the Walker phasing rule, given the reference phase $\boldsymbol{x}$, the phase coordinate of the $j_s$-th satellite in the $i_p$-th orbital plane is given by
\begin{equation}\label{eq:reference_coordinate}
    (\Omega_{i_p}, u_{i_p j_s}) = \boldsymbol{x} +
    \bigl(\tfrac{2\pi i_p}{N_o},\ \tfrac{2\pi j_s}{N_s} + \tfrac{2\pi F i_p}{N}\bigr)\pmod{2\pi},
\end{equation}
where $i_p \in \{0,1,\dots,N_o-1\}$ and $j_s \in \{0,1,\dots,N_s-1\}$ are the orbital plane index and the in-plane satellite index, respectively.

\subsubsection{Constellation Dynamics in the Phase Torus}

The \textit{constellation state} is defined as the set of phase coordinates of the $N$ satellites on the phase torus:
\begin{equation}\label{eq:constellation_state}
S(\boldsymbol{x}) = \bigl\{(\Omega_{i_p},\, u_{i_p j_s})\bigr\}_{i_p,\,j_s}.
\end{equation}
Equation~\eqref{eq:reference_coordinate} shows that, for fixed constellation parameters $(N_o, N_s, h, i, F)$, $S(\boldsymbol{x})$ is a rigid regular lattice whose state and temporal evolution are determined solely by the global translation $\boldsymbol{x}$.

Let $t$ denote time and $\boldsymbol{x}(0) = (\Omega_x(0), u_x(0))$ the \textit{initial phase} of the constellation.
The Earth's rotation reduces the LAN at the Earth rotation rate $\omega_E$, while the orbital revolution advances the in-plane phase at $\omega_s = 2\pi/T_{\mathrm{orb}}$, where $T_{\mathrm{orb}}$ is the orbital period.
The reference phase evolves as
\begin{equation}\label{eq:phase_flow}
\boldsymbol{x}(t) = \boldsymbol{x}(0) + \boldsymbol{v}\,t \pmod{2\pi}.
\end{equation}
Equation~\eqref{eq:phase_flow} shows that the reference phase undergoes a uniform linear flow on the phase torus, with velocity vector $\boldsymbol{v}=(-\omega_E,\omega_s)$.
Let $\rho\triangleq\omega_s/\omega_E$ denote the angular velocity ratio.

\begin{remark}
    Only the two rates $\omega_E$ and $\omega_s$, and hence their ratio $\rho$, enter the analysis. Orbital perturbations can therefore be absorbed by replacing $(-\omega_E,\omega_s)$ with the effective ECEF nodal and in-plane angular rates, provided that these rates remain constant over the analysis interval. RGT constellations are understood as idealizations with exactly rational $\rho$.
\end{remark}

\subsubsection{Support Set and Steady-State Distribution of the Reference Phase}

The \textit{support set} $\mathcal{C}\subseteq\mathbb{T}^2$ of the reference phase is the closure $\overline{\{\boldsymbol{x}(t) : t \ge 0\}}$ of the trajectory.
The geometry of $\mathcal{C}$ is determined by $\rho$ \cite{CB_WPP_1}, \cite[Sec.~1.5]{dynamical_theory}, and so is the measure $|\mathcal{C}|$, i.e., the area in $\mathbb{T}^2$ or the arc length along a geodesic. Two cases arise.

For a non-RGT constellation (irrational $\rho$), the trajectory is dense in the torus, so that $\mathcal{C} = \mathbb{T}^2$ and $|\mathcal{C}| = 4\pi^2$.
For an RGT constellation ($\rho = p/q$ with coprime integers $p$ and $q$), the trajectory is a closed geodesic that winds $q$ times in $\Omega$ and $p$ times in $u$ per period.
Since $\tfrac{d}{dt}(p\Omega+qu) = -p\omega_E + q\omega_s = 0$ under $\rho = p/q$, the invariant $c(\Omega,u)=p\Omega+qu \pmod{2\pi}$ labels the closed geodesic on which a satellite at $(\Omega,u)$ moves. The geodesic is the level set
\begin{equation}\label{eq:geodesic}
    \mathcal{C} = \{(\Omega, u)\in\mathbb{T}^2 : c(\Omega,u) = c_0\},
\end{equation}
where $c_0=c(\Omega_x,u_x)$ is the value at the reference satellite.
Consequently, the steady-state distribution of the reference phase depends on the initial phase only through the invariant $c_0$ in~\eqref{eq:geodesic} in the RGT case, and not at all in the non-RGT case \cite{CB_WPP_1}.

In steady state, the empirical distribution of $\boldsymbol{x}$ converges to the uniform probability measure $\mu_{\mathcal C}$ on $\mathcal{C}$. The limit follows from unique ergodicity of the irrational linear flow in the non-RGT case \cite{weyl1916gleichverteilung}, \cite[Prop.~4.2.1]{dynamical_theory}, and from periodic averaging along the closed geodesic in the RGT case.
Hence $\boldsymbol{x}$ can be treated as a random variable with distribution $\mu_{\mathcal C}$ \cite{CB_WPP_1}.

Accordingly, for any function $Z$ of the reference phase $\boldsymbol{x}$ that is $\mu_{\mathcal C}$-integrable and continuous $\mu_{\mathcal C}$-almost everywhere, the long-time average of $Z(\boldsymbol{x}(t))$ equals its statistical average under $\mu_{\mathcal C}$:
\begin{equation}\label{eq:steady_state_average}
    \lim_{T\to\infty}\frac{1}{T}\int_0^{T}Z(\boldsymbol{x}(t))\,\mathrm{d}t=\int_{\mathcal C}Z(\boldsymbol{x})\,\mathrm{d}\mu_{\mathcal C}(\boldsymbol{x}).
\end{equation}

\subsection{Receiver--Satellite Visibility on the Phase Torus}
\label{sec:visibility}

\subsubsection{Visibility Analysis}

Consider a receiver at geographic coordinates $(\lambda_{\mathrm{rx}}, \varphi_{\mathrm{rx}})$, where $\lambda_{\mathrm{rx}} \in [0, 2\pi)$ is the east longitude and $\varphi_{\mathrm{rx}} \in [-\pi/2, \pi/2]$ is the latitude. 
Let $\theta_{\min} \in [0, \pi/2)$ denote the minimum operational elevation angle, $\theta(\Omega,u)$ the elevation of a satellite at phase coordinate $(\Omega,u)$, and $\gamma(\Omega,u)$ the corresponding central angle at the Earth's center.

With $\hat{\boldsymbol{r}}_{\mathrm{rx}} = \bigl(\cos\varphi_{\mathrm{rx}}\cos\lambda_{\mathrm{rx}},\ \cos\varphi_{\mathrm{rx}}\sin\lambda_{\mathrm{rx}},\ \sin\varphi_{\mathrm{rx}}\bigr)^{\mathsf{T}}$ and $\hat{\boldsymbol{r}}_{\mathrm{sat}}$ from~\eqref{coordinates}, the central angle is obtained directly from the phase coordinate as
\begin{equation}\label{eq:gamma_def}
    \cos\gamma(\Omega,u) = \hat{\boldsymbol{r}}_{\mathrm{rx}}^{\mathsf{T}}\,\hat{\boldsymbol{r}}_{\mathrm{sat}}(\Omega,u) ,
\end{equation}
and the elevation follows from the central angle by elementary trigonometry,
\begin{equation}\label{eq:theta_of_gamma}
    \sin\theta(\Omega,u) = \frac{r\cos\gamma(\Omega,u) - R_e}{\sqrt{r^2 + R_e^2 - 2rR_e\cos\gamma(\Omega,u)}} .
\end{equation}
The elevation decreases monotonically with $\gamma(\Omega,u)$.

\begin{lemma}\label{lem:visibility}
Let $\boldsymbol{n}(\Omega) = (\sin i\sin\Omega,\, -\sin i\cos\Omega,\, \cos i)^{\mathsf{T}}$ be the orbit normal of the plane with LAN $\Omega$, $\beta(\Omega)$ the angle between $\hat{\boldsymbol{r}}_{\mathrm{rx}}$ and $\boldsymbol{n}(\Omega)$, and $\phi_0(\Omega)$ the in-plane phase minimizing the central angle to the receiver, well-defined whenever $\sin\beta(\Omega)>0$.
Let $\gamma_{\max} = \arccos\!\bigl(\frac{R_e}{r}\cos\theta_{\min}\bigr) - \theta_{\min}$ be the maximum visible central angle, and let
\begin{equation}\label{eq:interval_halfwidth}
    \Psi(\Omega) = \arccos\!\left(\frac{\cos\gamma_{\max}}{\sin\beta(\Omega)}\right) \in \left[0, \tfrac{\pi}{2}\right)
\end{equation}
be the half-width of the in-plane visibility interval, with $\Psi(\Omega)=0$ whenever $\sin\beta(\Omega)<\cos\gamma_{\max}$.
Then a satellite at phase coordinate $(\Omega,u)$ is visible to the receiver if and only if
\begin{equation}\label{eq:visibility_criterion}
    \begin{gathered}
        \theta(\Omega,u)\ge\theta_{\min}
        \iff \cos{\gamma(\Omega,u)}\ge\cos{\gamma_{\max}}\\
        \iff
        \begin{cases}
            \beta \in \bigl[\tfrac{\pi}{2} - \gamma_{\max},\ \tfrac{\pi}{2} + \gamma_{\max}\bigr],\\[1ex]
            u \in \bigl[\phi_0 - \Psi,\ \phi_0 + \Psi\bigr] \pmod{2\pi},
        \end{cases}
    \end{gathered}
\end{equation}
with $\beta$, $\phi_0$, and $\Psi$ evaluated at $\Omega$.
The proof is given in Appendix~\ref{app:lemma1}.
\end{lemma}

The receiver--satellite distance and the off-axis angle $\alpha$ at the satellite, measured from its nadir direction, are
\begin{equation}\label{eq:slant_range}
    \begin{gathered}
    d(\theta) = \sqrt{r^2 - R_e^2\cos^2\theta} - R_e\sin\theta,\\
    \sin\alpha = \frac{R_e}{r}\cos\theta .
    \end{gathered}
\end{equation}

\subsubsection{Visibility Window}
The \textit{visibility region} $\mathcal{V}_{\mathrm{rx}}$ of receiver $\mathrm{rx}$ is the set of points on the phase torus satisfying the visibility condition, and its indicator function is termed the \textit{visibility window}:
\begin{equation}\label{eq:visibility_window}
    \begin{gathered}
    \mathcal{V}_{\mathrm{rx}} = \bigl\{(\Omega, u)\in\mathbb{T}^2 : \cos\gamma(\Omega, u) \ge \cos\gamma_{\max}\bigr\},\\
    V_{\mathrm{rx}}(\Omega, u) = \mathds{1}\bigl\{(\Omega, u)\in\mathcal{V}_{\mathrm{rx}}\bigr\}.
    \end{gathered}
\end{equation}
The receiver longitude merely translates $\mathcal{V}_{\mathrm{rx}}$ along the $\Omega$-direction without changing its shape, whereas the latitude alters both the shape and the position along the $u$-direction \cite{Quick2Dmap}.

Moreover, by the visibility condition in Lemma~\ref{lem:visibility}, the 2-D Fourier transform of the visibility window reduces to a 1-D integral:
\begin{equation}\label{eq:vis_window_ft}
    \begin{aligned}
        \hat{V}_{\mathrm{rx}}(a,b) &= \iint_{\mathbb{T}^2} V_{\mathrm{rx}}(\Omega,u)\,e^{-j(a\Omega+bu)}\,\mathrm{d}\Omega\,\mathrm{d}u \\
        &= 2\int_0^{2\pi} e^{-ja\Omega-jb\phi_0(\Omega)}\,\frac{\sin\bigl(b\Psi(\Omega)\bigr)}{b}\,\mathrm{d}\Omega,
    \end{aligned}
\end{equation}
where $\sin(b\Psi)/b$ is read as $\Psi$ at $b=0$, and the convention $\Psi(\Omega)=0$ for $\sin\beta(\Omega)<\cos\gamma_{\max}$ makes the integrand vanish for orientations with no visible arc, so the integral over $[0,2\pi]$ remains exact.

\begin{figure*}[!t]
    \centering
    \begin{minipage}[t]{0.49\textwidth}
        \centering
        \subfloat[The receiver--satellite visibility in 3-D space\label{fig:vis-a}]{\includegraphics{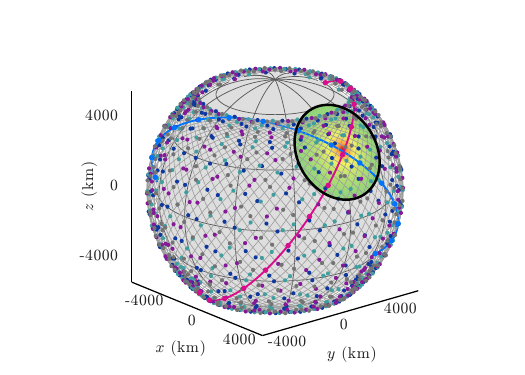}}
    \end{minipage}\hfill
    \begin{minipage}[t]{0.49\textwidth}
        \centering
        \subfloat[The receiver--satellite visibility on the phase torus\label{fig:vis-b}]{\includegraphics{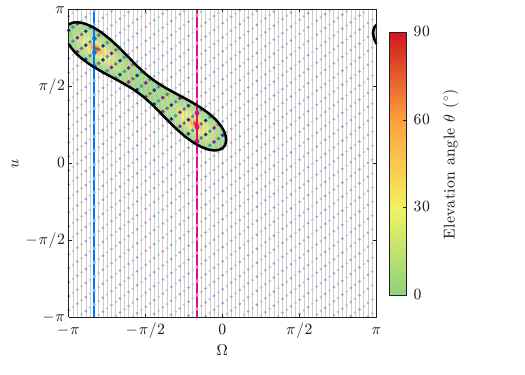}}
    \end{minipage}
    \caption{Receiver--satellite visibility for the Starlink Gen1 sub-constellation $53^\circ\!:\!1584/72/17$ ($h=550$~km) at $35^\circ$N with $\theta_{\min}=0^\circ$.}
    \label{fig:visibility_window}
\end{figure*}

Fig.~\ref{fig:visibility_window} shows the same visibility region in two coordinates.
In ECEF coordinates (Fig.~\ref{fig:visibility_window}a), the region is a spherical cap on the orbit shell. On the phase torus (Fig.~\ref{fig:visibility_window}b), the region is a band, with the color encoding elevation in both panels.
The two highlighted orbital planes in Fig.~\ref{fig:visibility_window}(a) cross the center of the receiver's FOV, one on an ascending pass and one on a descending pass. On the torus the two orbits appear as two vertical lines.
Because the window is $2\pi$-periodic in both coordinates, the band re-enters at the opposite edge and therefore appears cut at the boundary of Fig.~\ref{fig:visibility_window}(b).

\subsection{Observables}
\label{sec:observables}

The \textit{per-satellite observable} $f_{\mathrm{rx}}(s)$ is the contribution of a satellite at phase coordinate $s=(\Omega,u)$ to receiver $\mathrm{rx}$, assumed to be bounded, piecewise smooth, and scalar-valued on the phase torus, with discontinuities forming a finite union of smooth curves.

The \textit{additive observable} $Z_{\mathrm{rx}}(\boldsymbol{x})$ (hereinafter the observable) is the sum of $f_{\mathrm{rx}}(s)$ over all visible satellites for a given reference phase $\boldsymbol{x}$, i.e.,
\begin{equation}\label{eq:Zrx_def}
    Z_\mathrm{rx}(\boldsymbol{x}) \triangleq \sum_{\substack{s \in S(\boldsymbol{x}) \\ V_\mathrm{rx}(s) = 1}} f_\mathrm{rx}(s).
\end{equation}

Direct analysis of~\eqref{eq:Zrx_def} is hindered by two facts: the contributing set $S(\boldsymbol{x})\cap\mathcal{V}_{\mathrm{rx}}$ evolves with the constellation motion and can only be found by testing every satellite, and the start and end times of a single-satellite pass are roots of the transcendental equation $\cos\gamma(\Omega(t),u(t))=\cos\gamma_{\max}$, which couples the two angular frequencies and admits no closed form.

Rewriting the summation over the full constellation removes the first obstruction and absorbs the second into a fixed function of the phase coordinate:
\begin{align}
    Z_\mathrm{rx}(\boldsymbol{x})&=\sum_{s\in S(\boldsymbol{x})}f_\mathrm{rx}(s)\cdot V_\mathrm{rx}(s) \notag\\
    &=\sum_{s\in S(\boldsymbol{x})}U_{\mathrm{rx}}(s), \label {eq:additive_observable}
\end{align}
where $U_{\mathrm{rx}}(s) \triangleq f_{\mathrm{rx}}(s)V_{\mathrm{rx}}(s)$.
For a fixed receiver, $U_{\mathrm{rx}}$ depends solely on the satellite phase coordinate and is termed the receiver's \textit{observation field} (hereinafter the \textit{field}). At the same latitude, a common longitude rotation leaves the geometry invariant, so the field of a receiver at longitude $\lambda_{\mathrm{rx}}$ is the translate $U_{\mathrm{rx}}(\Omega,u)=U_{0}(\Omega-\lambda_{\mathrm{rx}},u)$ of the field $U_{0}$ at $\lambda_{\mathrm{rx}}=0$.
By construction, the field is bounded and piecewise smooth on $\mathbb{T}^2$, with discontinuities confined to the boundary of the visibility window and the curves along which $f_{\mathrm{rx}}$ jumps, and no further restriction is imposed on the functional form of $f_{\mathrm{rx}}$.

%% file: sections/performance_analysis.tex
%% ============================================================
%%  Section III: Performance Analysis —— 一般框架
%%  对应中文稿：第 III 章「性能分析」。
%%  第 IV 章「下行集总干扰」的应用（含数值验证）见 sections/interference.tex。
%% ============================================================
\section{Steady-state Performance Analysis of Observables}
\label{sec:performance}

%% ---------------- 一般理论 ---------------- %%

\subsection{Spectral Representation of Observables}
\label{sec:spec-rep}
\begin{definition}[Phase lattice and dual lattice]\label{def:lattices}
    The \textit{phase lattice} $\mathcal{L}$ of the constellation is defined as the lattice generated by
    \begin{equation}\label{eq:phase_lattice}
        \mathcal{L}=\bigl\{i_p\boldsymbol{x}_1 + j_s\boldsymbol{x}_2 : i_p, j_s \in \mathbb{Z}\bigr\},
    \end{equation}
    where $\boldsymbol{x}_1 = (\tfrac{2\pi}{N_o},\, F\tfrac{2\pi}{N})$ and $\boldsymbol{x}_2 = (0,\, \tfrac{2\pi}{N_s})$.
    The phase lattice $\mathcal{L}$ depends only on the constellation parameters $(N_o, N_s, F)$.
    Its intersection with the phase torus $\mathbb{T}^2$ (in the mod-$2\pi$ sense) is exactly the set of satellite phase coordinates at $\boldsymbol{x}=\boldsymbol{0}$.
    The \textit{dual lattice} of $\mathcal{L}$ is
    \begin{equation}\label{eq:dual_lattice}
        \mathcal{L}^\perp=\{(mN_o - Fn,\; nN_s) : m,n\in\mathbb{Z}\}.
    \end{equation}
\end{definition}

\begin{theorem}[Spectral representation]\label{theo:spectral_rep}
    Represent the constellation by the impulse-array measure $\Delta_{\boldsymbol{x}}\triangleq\sum_{\boldsymbol{s}\in S(\boldsymbol{x})}\delta_{\boldsymbol{s}}$, so that $Z_{\mathrm{rx}}(\boldsymbol{x})=\int_{\mathbb{T}^2}U_{\mathrm{rx}}\,\mathrm{d}\Delta_{\boldsymbol{x}}$. The additive observable admits the spectral representation
    \begin{equation}\label{eq:spectral_rep}
        Z_{\mathrm{rx}}(\boldsymbol{x})=\zeta\sum_{\boldsymbol{y}\in\mathcal{L}^\perp}\hat{U}_{\mathrm{rx}}(\boldsymbol{y})\,e^{j\boldsymbol{y}^{\mathsf{T}}\boldsymbol{x}},
        \qquad \zeta=\frac{N}{4\pi^2},
    \end{equation}
    where $\hat{U}_{\mathrm{rx}}$ denotes the 2-D Fourier transform of $U_{\mathrm{rx}}$. The series converges to $Z_{\mathrm{rx}}$ in $L^2(\mathbb{T}^2)$. The proof is given in Appendix~\ref{app:theo1}.
\end{theorem}

The constant $\zeta$ is the satellite density on the phase torus, i.e., the expected number of satellites per unit torus area.
Substituting the phase flow~\eqref{eq:phase_flow} into~\eqref{eq:spectral_rep} yields an explicit spectral representation of the temporal evolution of the observable:
\begin{equation}\label{eq:spectral_rep_with_t}
    Z_\mathrm{rx}(t)=\zeta\sum_{\boldsymbol{y}\in\mathcal{L}^\perp}\hat{U}_{\mathrm{rx}}(\boldsymbol{y})e^{j\boldsymbol{y}^{\mathsf{T}}\boldsymbol{x}(0)}e^{j\boldsymbol{y}^{\mathsf{T}}\boldsymbol{v}t}
\end{equation}
Hence $Z_{\mathrm{rx}}(t)$ is a superposition of constant-envelope complex exponentials at frequencies $\boldsymbol{y}^{\mathsf{T}}\boldsymbol{v}$, with amplitudes given by the field spectrum $\hat{U}_{\mathrm{rx}}$ sampled on the dual lattice $\mathcal{L}^\perp$.

The three factors shaping the observable are thereby decoupled: the constellation configuration is captured by the dual lattice $\mathcal{L}^\perp$; the receiver geometry and the functional form of the per-satellite observable are encoded in the field spectral coefficients $\hat{U}_{\mathrm{rx}}(\boldsymbol{y})$; and the constellation state enters through the reference phase $\boldsymbol{x}$, which further separates into the initial phase $\boldsymbol{x}(0)$ and the temporal evolution $\boldsymbol{v}t$, carried respectively by $e^{j\boldsymbol{y}^{\mathsf{T}}\boldsymbol{x}(0)}$ and $e^{j\boldsymbol{y}^{\mathsf{T}}\boldsymbol{v}t}$.

\begin{figure}[htbp]
    \centering
    \includegraphics[width=\columnwidth]{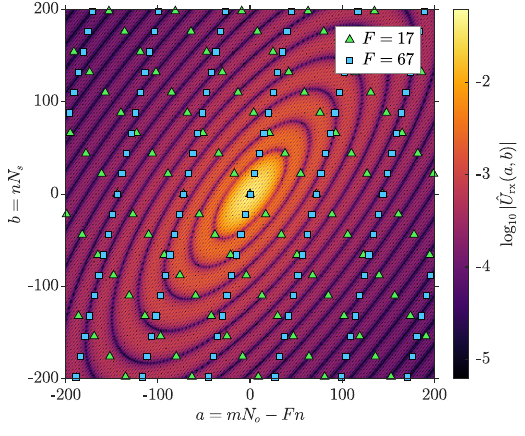}
    \caption{Separation of the constellation configuration from the observation field in the spectral representation of the aggregate downlink interference.}
    \label{fig:spectrum_and_lattice}
\end{figure}

Fig.~\ref{fig:spectrum_and_lattice} illustrates this separation for the downlink aggregate interference, using the Starlink Gen1 sub-constellation with $\theta_{\min}=10^\circ$, a receiver at $35^\circ$N, and the link parameters of the non-RGT example in Section~\ref{sec:intf-validation}.
The background shows the field spectrum $\log_{10}\lvert\hat{U}_{\mathrm{rx}}(a,b)\rvert$ over the spatial frequency window $\lvert a\rvert,\lvert b\rvert\le200$, while the green triangles and blue squares mark the dual lattices sampled under the phasing factors $F=17$ and $F=67$, respectively.
Since $\mathcal{L}^\perp$ depends only on $(N_o,N_s,F)$, changing the configuration moves only the foreground lattice points at which the spectrum is sampled, and the ring-shaped background remains fixed.
The phasing factor shears the phase lattice along the $u$-direction and, by duality, the dual lattice along the $a$-direction.

In particular, for RGT constellations the reference phase is confined to the closed geodesic $\mathcal{C}\subset\mathbb{T}^2$. Substituting the parameterization of $\mathcal{C}$ into~\eqref{eq:spectral_rep_with_t} and collecting the terms that share the same temporal frequency yields a Fourier series in time, as stated in Theorem~\ref{theo:spectral_rep_recurr}.

\begin{theorem}[Spectral representation in RGT constellations]\label{theo:spectral_rep_recurr}
    For an RGT constellation, the following two statements hold.
    \begin{enumerate}
        \item[(a)] \textit{Representation.} The observable evolves in time as
        \begin{equation}\label{eq:spectral_rep_recur_t}
            Z_\mathrm{rx}(t) = \zeta \sum_{l \in g_t \mathbb{Z}} \hat{H}_l \, e^{j l \omega_E t / q},
        \end{equation}
        a discrete Fourier series in time whose harmonic index $l$ runs over the multiples of $g_t=\gcd(pN_s+qF,\,qN_o)$, with the coefficients
        \begin{equation}\label{eq:Hl}
            \hat{H}_l = \sum_{\substack{\boldsymbol{y} = (a,b) \in \mathcal{L}^\perp \\ pb - qa = l}}
            \hat{U}_0(a,b)\, e^{j\left(a\Omega_x(0) + b u_x(0) - a\lambda_{\mathrm{rx}}\right)}
        \end{equation}
        Here $\hat U_0$ is the Fourier transform of $U_0$, with $\hat U_{\mathrm{rx}}(a,b)=e^{-ja\lambda_{\mathrm{rx}}}\hat U_0(a,b)$; the sum collects all dual-lattice points that share the same temporal frequency. The equality holds in $L^2$ over one trajectory period.
        \item[(b)] \textit{Well-posedness.} If $\lvert\hat U_0(\boldsymbol{y})\rvert=O(\lvert\boldsymbol{y}\rvert^{-3/2})$ on $\mathcal{L}^\perp$, each sum in~\eqref{eq:Hl} converges absolutely, so that every coefficient is well defined.
    \end{enumerate}
    The proof is given in Appendix~\ref{app:theo2}.
\end{theorem}

\begin{remark}\label{rem:temporal_period}
    The temporal frequencies in~\eqref{eq:spectral_rep_with_t} are the values $\boldsymbol{y}^{\mathsf{T}}\boldsymbol{v}$ taken over the dual lattice.
    Writing $\boldsymbol{y}=(mN_o-Fn,\,nN_s)$,
    \begin{equation*}
        \boldsymbol{y}^{\mathsf{T}}\boldsymbol{v}
        = \omega_E\bigl[-mN_o+n\,(F+N_s \rho)\bigr],
    \end{equation*}
    so the frequencies form the subgroup of $\mathbb{R}$ generated by $-N_o\omega_E$ and $(F+N_s\rho)\omega_E$.
    For irrational $\rho$ (non-RGT), this subgroup is dense; a minimum positive period can then exist only when the active coefficients degenerate to a discrete frequency group. For rational $\rho$ (RGT), the subgroup is discrete and reduces to $\{l\omega_E/q\}_{l\in g_t\mathbb{Z}}$.
    The observable then has period $2\pi q/(g_t\omega_E)$. For a nonconstant observable, its minimum positive period is $2\pi q/(G_t\omega_E)$, where $G_t=\gcd\{l:\hat{H}_l\neq0\}$ is a multiple of $g_t$. In particular, the minimum equals $2\pi q/(g_t\omega_E)$ whenever at least one of $\hat H_{\pm g_t}$ is nonzero.
    Hence the performance period is $T_{\mathrm{per}}=T_{\mathrm{geo}}/G_t$, rather than the ground-track repeat period $T_{\mathrm{geo}}=2\pi q/\omega_E$, and is at most $T_{\mathrm{geo}}/g_t$.
\end{remark}

\subsection{Longitude Periodicity of Walker-Constellation Observables}
\label{sec:longitude-period}

According to Theorem~\ref{theo:spectral_rep}, the receiver's position is encoded solely in the field spectrum. Hence, the longitude dependence of the observable can be analyzed separately.

\begin{theorem}[Instantaneous longitude period of observables in general constellations]\label{theo:inst_period}
    For any Walker constellation, fix the time instant $t$ and suppose that at least one nonzero dual-lattice coefficient has $a\neq0$. The minimum positive period of $Z_{\mathrm{rx}}(t)$ with respect to the receiver longitude $\lambda_{\mathrm{rx}}$ is then
    \begin{equation}
        \frac{2\pi}{G_\lambda},
        \qquad
        G_\lambda=\gcd\{a:\exists b,\ (a,b)\in\mathcal{L}^\perp,\ \hat U_0(a,b)\neq0\}.
    \end{equation}
    If no such coefficient exists, the observable is independent of longitude (possibly identically zero), and no minimum positive longitude period is defined. When at least one dual-lattice coefficient with $a=\pm\gcd(N_o,F)$ is nonzero, $G_\lambda=\gcd(N_o,F)$, and the period is $\Delta\lambda_{\mathrm{inst}}=2\pi/\gcd(N_o,F)$.
    The detailed proof is given in Appendix~\ref{app:theo3}.
\end{theorem}

\begin{remark}\label{rem:inst_period}
    In the nondegenerate case, for any constellation state and any time instant, the observable at $\mathrm{rx}=(\lambda_{\mathrm{rx}},\varphi_{\mathrm{rx}})$ is identical to that at $\mathrm{rx}'=(\lambda_{\mathrm{rx}}+k\Delta\lambda_{\mathrm{inst}},\varphi_{\mathrm{rx}})$, $k\in\mathbb Z$, i.e., $Z_{\mathrm{rx}}(t)\equiv Z_{\mathrm{rx}'}(t)$.
    When $N_o$ and $F$ are coprime, $\Delta\lambda_{\mathrm{inst}}=2\pi$.
    No two locations at the same latitude then share the same performance at all time instants.
\end{remark}

\begin{theorem}[Steady-state longitude period of observables in RGT constellations]\label{theo:ss_period}
    For an RGT Walker constellation, the steady-state distribution of an observable has the longitude period
    \begin{equation}
        \Delta\lambda_{\mathrm{steady}}
        =
        \frac{2\pi}{pN}\, g_c, \quad
        g_c = \gcd\bigl(pN_s+qF,\ qN_o,\ N\bigr).
    \end{equation}
    This is the minimum positive period when the observable distinguishes distinct translates of the trajectory set. Steady-state moments share every longitude period of the distribution, although their minimum period can be smaller.
    The detailed proof is given in the Appendix~\ref{app:theo4}.
\end{theorem}

\begin{corollary}\label{cor:period_ratio}
    For an RGT Walker constellation with at least one nonzero dual-lattice coefficient satisfying
    $a=\pm\gcd(N_o,F)$, the ratio of the instantaneous period of
    Theorem~\ref{theo:inst_period} to the steady-state period of
    Theorem~\ref{theo:ss_period} is a positive integer; explicitly,
    \begin{equation}
        \frac{\Delta\lambda_{\mathrm{inst}}}{\Delta\lambda_{\mathrm{steady}}}
        = \frac{pN}{g_c\gcd(N_o,F)} \in \mathbb{Z}^{+}.
    \end{equation}
    The proof is given in Appendix~\ref{app:cor1}.
\end{corollary}

For non-RGT constellations, the trajectory is dense on the torus and no periodic spacing exists. The reference phase then sweeps $\mathbb{T}^2$ uniformly, and, since the uniform measure on the torus is invariant under a translation along $\Omega$, the steady-state performance is independent of the receiver longitude.

\subsection{Analytical Expressions for Moments}
\label{sec:moments}
\begin{theorem}[Expectation and variance of the observable in non-RGT constellations]\label{theo:moments_for_non_recurr}
    For a non-RGT constellation, the steady-state expectation and variance of the observable are
    \begin{equation}\label{eq:moments}
        \mathbb{E}[Z_\mathrm{rx}] = \zeta\,\hat{U}_{\mathrm{rx}}(\boldsymbol{0}), \qquad
        \mathrm{Var}(Z_\mathrm{rx}) = \zeta^2 \sum_{\boldsymbol y\in\mathcal{L}^\perp\setminus\{\boldsymbol{0}\}} \bigl|\hat{U}_{\mathrm{rx}}(\boldsymbol{y})\bigr|^2 .
    \end{equation}
    The proof is given in Appendix~\ref{app:theo5}.
\end{theorem}

The coefficient at $\boldsymbol{y}=\boldsymbol{0}$ gives the expectation, $\mathbb{E}[Z_\mathrm{rx}]=\zeta\,\hat{U}_{\mathrm{rx}}(\boldsymbol{0})=N\langle U_{\mathrm{rx}}\rangle$, where $\langle U_{\mathrm{rx}}\rangle=\frac{1}{4\pi^2}\iint_{\mathbb{T}^2}U_{\mathrm{rx}}\,\mathrm{d}\boldsymbol{x}$ is the torus-averaged field. 
The expectation thus depends on the constellation only through its size $N$ and is independent of detailed configurations $(N_o,N_s,F)$.
For the variance, Parseval's identity gives $(4\pi^2)^{-1}\sum_{\boldsymbol{y}}\lvert\hat{U}_{\mathrm{rx}}(\boldsymbol{y})\rvert^2=\iint_{\mathbb{T}^2}\lvert U_{\mathrm{rx}}(\boldsymbol{x})\rvert^2\,\mathrm{d}\boldsymbol{x}<\infty$, so the series in~\eqref{eq:moments} is summable and equals the spectral energy carried by the dual-lattice points $\boldsymbol{y}\neq\boldsymbol{0}$.

Numerical evaluation of the variance via~\eqref{eq:moments} requires summing an infinite series of nonnegative terms, so truncation can only underestimate the true variance.

\begin{theorem}[Expectation and variance of the observable in RGT constellations]\label{theo:moments_for_recurr}
    For an RGT constellation, with the reference phase $\boldsymbol{x}$ uniformly distributed on the closed geodesic $\mathcal{C}$ of~\eqref{eq:geodesic}, the steady-state expectation and variance of the observable are
    \begin{equation}\label{eq:moments_for_recurr}
        \mathbb{E}[Z_\mathrm{rx}] = \zeta\,\hat{H}_0, \qquad
        \mathrm{Var}(Z_\mathrm{rx}) = \zeta^2 \sum_{l\in g_t\mathbb{Z}\setminus\{0\}} \bigl|\hat{H}_l\bigr|^2 .
    \end{equation}
\end{theorem}
\begin{IEEEproof}
    Since the exponentials $e^{jl\omega_E t/q}$ are orthogonal over one trajectory period, only the $l=0$ term survives in the mean. Expanding $\lvert Z_{\mathrm{rx}}-\mathbb{E}Z_{\mathrm{rx}}\rvert^2$ and using the same orthogonality eliminate all cross terms, giving $\mathbb{E}[Z_\mathrm{rx}]=\zeta\hat{H}_0$ and $\mathrm{Var}(Z_\mathrm{rx})=\zeta^2\sum_{l\in g_t\mathbb{Z}\setminus\{0\}}\lvert\hat{H}_l\rvert^2$.
\end{IEEEproof}

The RGT variance~\eqref{eq:moments_for_recurr} is not the restriction of~\eqref{eq:moments} to the geodesic, because the two average the observable over different sets. Averaging over the whole phase torus erases all dependence on the receiver longitude and the initial phase, whereas averaging along a single closed geodesic retains both.

\subsection{Steady-State Distribution}
\label{sec:ss-dist}
The spectral, periodicity, and moment results above hold for complex-valued observables.
The distributional analysis below requires $Z_{\mathrm{rx}}$ to be real, whose characteristic function (CF) is
\begin{equation}\label{eq:CF_def}
    \Phi_Z(\xi) \triangleq \mathbb{E}\left[e^{j\xi Z_\mathrm{rx}}\right] .
\end{equation}
By the definition of the steady-state distribution on the phase torus given in \eqref{eq:steady_state_average}, the CF is evaluated directly over the support set $\mathcal{C}$ as
\begin{equation}\label{eq:CF_compute}
    \Phi_Z(\xi) = \int_{\mathcal C} e^{j\xi Z_\mathrm{rx}(\boldsymbol{x})}\,\mathrm{d}\mu_{\mathcal C}(\boldsymbol{x}).
\end{equation}

The steady-state distribution is reported through its CCDF,
\begin{equation}\label{eq:CCDF_inversion}
    \bar F_Z(z) \triangleq \Pr(Z_{\mathrm{rx}}>z)
    = \frac12+\frac1\pi\int_0^\infty\frac{\operatorname{Im}\left[e^{-j\xi z}\Phi_Z(\xi)\right]}{\xi}\,\mathrm{d}\xi ,
\end{equation}
obtained from the CF by the Lévy inversion, with the value at a jump taken as the average of its left and right limits.

For a lattice-valued observable with equally spaced values, $Z_{\mathrm{rx}}\in\{z_k = z_0+k\Delta z\}_{k\in\mathcal{K}}$ with $\mathcal{K}\subseteq\mathbb{Z}$ and $\Delta z>0$, the same formulas apply and $\bar F_Z$ is a step function.
Additivity bounds the observable by at most $N$ contributions of a bounded field, so $\mathcal{K}$ is finite and $\Phi_Z$ is a trigonometric polynomial of period $2\pi/\Delta z$. 
Sampling the CF at $M\ge\max\mathcal{K}-\min\mathcal{K}+1$ equispaced frequencies then recovers the jumps exactly,
\begin{equation}\label{eq:PMF_inversion}
    p_k = \frac{1}{M}\sum_{m=0}^{M-1}\Phi_Z(\xi_m)\,e^{-j\xi_m z_k},
    \quad \xi_m = \frac{2\pi m}{M\Delta z},\; k\in\mathcal{K},
\end{equation}
an exact algebraic identity, free of the truncation error that \eqref{eq:CCDF_inversion} incurs in practice.

%% file: sections/interference.tex
%% ============================================================
%%  Section IV: Aggregate Downlink Interference
%%  —— 一般框架在下行集总干扰上的完整应用（解析 + 数值验证）
%%  A 模型与谱表示 · B 矩与稳态分布 · C 仰角分辨结果与强度密度
%%  D 数值验证 · E SG 基线偏差机理
%%  对应中文稿：第 IV 章（IV-A/IV-B/IV-C 解析 + IV-D 数值验证）
%%  说明：一般理论（谱表示、经度周期、矩、稳态分布）见 sections/performance_analysis.tex
%% ============================================================
\section{Aggregate Downlink Interference Analysis and Validation}
\label{sec:interference}
\subsection{Interference Model and Spectral Representation}
\label{sec:intf-model}

We now take the per-satellite observable to be the power received by a receiver from a single satellite, so that the additive observable becomes the aggregate downlink interference power.

All satellites are identical, transmitting with the same power $P_s$ while pointing their antenna boresights to the nadir.
The transmit and receive antenna gains are $G_{\mathrm{t}}$ and $G_{\mathrm{r}}$, the receiver antenna being assumed to have a time-invariant pointing, and the carrier wavelength is $\lambda_c$. With the receiver--satellite distance $d(\theta)$ of Section~\ref{sec:visibility}, the power received from a single satellite defines a per-satellite observable,
\begin{equation}\label{eq:f_rx_I}
    f_{\mathrm{rx},I}(\Omega,u) = P_s\left(\frac{\lambda_c}{4\pi d(\theta)}\right)^{2} G_{\mathrm{r}}\,G_{\mathrm{t}}.
\end{equation}
The aggregate downlink interference power $I_{\mathrm{rx}}$ is the additive observable,
\begin{equation}\label{eq:Irx_def}
    I_{\mathrm{rx}}(\boldsymbol{x}) = \sum_{s\in S(\boldsymbol{x})} f_{\mathrm{rx},I}(s)\,V_{\mathrm{rx}}(s)
    = \sum_{s\in S(\boldsymbol{x})} U_{\mathrm{rx},I}(s) ,
\end{equation}
where $U_{\mathrm{rx},I} \triangleq V_{\mathrm{rx}} f_{\mathrm{rx},I}$ is the interference observation field.

Since $f_{\mathrm{rx},I}$ and the visibility window are bounded and piecewise smooth, with jump discontinuities along smooth level sets, their product $U_{\mathrm{rx},I}$ meets the standing assumption of Section~\ref{sec:observables}, so that the framework applies to $I_{\mathrm{rx}}$ verbatim without assuming that satellite positions are conditionally independent. 
In particular, the spectral representation of the interference follows from Theorem~\ref{theo:spectral_rep} as
\begin{equation}\label{eq:I_spectral}
    I_{\mathrm{rx}}(\boldsymbol{x}) = \zeta\sum_{\boldsymbol{y}\in\mathcal{L}^\perp}\hat{U}_{\mathrm{rx},I}(\boldsymbol{y})\,e^{j\boldsymbol{y}^{\mathsf{T}}\boldsymbol{x}} .
\end{equation}

\subsection{Moments and Steady-State Distribution}
\label{sec:intf-moments}
For the non-RGT case, the steady-state expectation and variance of the interference are obtained by substituting $U_{\mathrm{rx},I}$ into Theorem~\ref{theo:moments_for_non_recurr}:
\begin{equation}
    \mathbb{E}[I] = \zeta\,\hat{U}_{\mathrm{rx},I}(\boldsymbol{0}), \quad
    \mathrm{Var}(I) = \zeta^2 \sum_{\mathcal{L}^\perp\setminus\{\boldsymbol{0}\}} \bigl|\hat{U}_{\mathrm{rx},I}(\boldsymbol{y})\bigr|^2.
\end{equation}
For the RGT case, the moments are given by Theorem~\ref{theo:moments_for_recurr}: $\mathbb{E}[I]=\zeta\hat H_0$ and $\mathrm{Var}(I)=\zeta^2\sum_{l\neq0}|\hat H_l|^2$, where $\hat H_l$ is obtained by substituting $U_{\mathrm{rx},I}$ into~\eqref{eq:Hl}.

Since $I_{\mathrm{rx}}$ is a real-valued observable, its steady-state CCDF follows from the support-set average of the CF, \eqref{eq:CF_compute}, through the Lévy inversion \eqref{eq:CCDF_inversion}:
\begin{equation}\label{eq:I_ccdf}
    \begin{aligned}
        \Phi_I(\xi) &= \int_{\mathcal C} e^{j\xi I(\boldsymbol{x})}\,\mathrm{d}\mu_{\mathcal C}(\boldsymbol{x}),\\
        \bar F_I(z) &= \frac12+\frac1\pi\int_0^\infty\frac{\operatorname{Im}\left[e^{-j\xi z}\Phi_I(\xi)\right]}{\xi}\,\mathrm{d}\xi .
    \end{aligned}
\end{equation}

The above conclusions do not rely on the specific form of the satellite and receiver antenna patterns.

\begin{remark}[Connection to EPFD assessment]\label{rem:epfd}
For a GSO earth station, specifying its location, pointing direction, and antenna pattern in the observation field makes $\bar F_I(z)$ the long-term CCDF of the aggregate interference contributed by a single NGSO constellation. After incorporating the regulatory reference bandwidth and antenna-gain normalization into $f_{\mathrm{rx},I}$, EPFD exceedance probabilities and percentile levels can be obtained directly from this CCDF without time-stepped statistics. This provides an analytical single-constellation building block for the EPFD assessment motivated in Section~\ref{sec:intro}. Aggregation over multiple constellations and the complete ITU-R compliance procedure \cite{itur2023s1503} are beyond the present scope.
\end{remark}

% \clearpage
\subsection{Numerical Validation}
\label{sec:intf-validation}

\begin{table}[!t]
    \centering
    \caption{Parameters of the Non-RGT Example and the RGT Example}
    \label{tab:sim-params}
    \footnotesize
    \setlength{\tabcolsep}{3pt}
    \begin{tabular}{lll}
        \hline
        Parameter & Non-RGT & RGT \\
        \hline
        \multicolumn{3}{l}{\textit{Constellation}} \\
        Orbital altitude $h$ & 550~km & 11090.38~km \\
        Planes $N_o$ / satellites $N_s$ & $72$ / $22$ & $6$ / $12$ \\
        Satellites $N$ / phasing $F$ & $1584$ / $17$ \cite{Starlink_phasing_factor_17} & $72$ / $3$ \\
        Walker-$\delta$ type & Non-RGT & RGT ($15\!:\!4$) \\
        Ground-track repeat period $T_{\mathrm{geo}}$ & -- & 4 sidereal days \\[4pt]
        \multicolumn{3}{l}{\textit{Link}} \\
        Transmit cap $\psi_{\mathrm{t}}$ & $45^\circ$ & $11^\circ$ \\[4pt]
        \multicolumn{3}{l}{\textit{Receiver}} \\
        Longitude $\lambda_{\mathrm{rx}}$ & $0^\circ$ & $0^\circ$--$120^\circ$ sweep \\
        CCDF latitudes & $0^\circ$, $35^\circ$, $60^\circ$ & $35^\circ$ \\
        Moment latitudes & $0^\circ$--$70^\circ$ & $35^\circ$ \\[4pt]
        \multicolumn{3}{l}{\textit{Simulation}} \\
        Duration / time step & 30 days / 30~s & 4 sidereal days / 1~s \\[4pt]
        \multicolumn{3}{l}{\textit{Common}} \\
        Orbital inclination $i$ & \multicolumn{2}{c}{$53^\circ$} \\
        Initial reference phase & \multicolumn{2}{c}{$0$} \\
        Minimum elevation $\theta_{\min}$ & \multicolumn{2}{c}{$10^\circ$} \\
        Gains $G_{\mathrm{h}}/G_{\mathrm{l}}$ & \multicolumn{2}{c}{$24$~dBi / $0$~dBi} \\
        Receiver gain $G_{\mathrm{r}}$ & \multicolumn{2}{c}{0~dBi} \\
        Path loss / fading & \multicolumn{2}{c}{Free space / neglected} \\
        Power normalization & \multicolumn{2}{c}{Zenith high-gain $=1$} \\
        Earth radius $R_e$ & \multicolumn{2}{c}{$6378.137$~km} \\
        Gravitational parameter $\mu$ & \multicolumn{2}{c}{$398600.4418$~km$^3$/s$^2$} \\
        Earth rotation rate $\omega_E$ & \multicolumn{2}{c}{$7.2921159\times10^{-5}$~rad/s} \\
        Dual-lattice cutoff & \multicolumn{2}{c}{$\sqrt{a^2+b^2}\le4800$} \\
        \hline
    \end{tabular}
\end{table}

\subsubsection{Non-RGT Setup}
\label{sec:exp-setup}
The study uses the non-RGT Starlink Gen1 sub-constellation, with the orbital parameters in Table~\ref{tab:sim-params}. Because the operational phasing factor is not public, $F=17$ is adopted \cite{Starlink_phasing_factor_17}. The transmit antenna follows a two-level cap pattern with half-angle $\psi_{\mathrm{t}}$,
\begin{equation}
    G_{\mathrm{t}}(\alpha) =
    \begin{cases}
        G_{\mathrm{h}}, & \alpha \le \psi_{\mathrm{t}},\\
        G_{\mathrm{l}}, & \alpha > \psi_{\mathrm{t}},
    \end{cases}
\end{equation}
so the observable is piecewise smooth with jumps only on the level set $\{\alpha(\Omega,u)=\psi_{\mathrm{t}}\}$ (a finite union of smooth closed curves for regular $\psi_{\mathrm{t}}$). The parameter values are listed in Table~\ref{tab:sim-params}, inspired by the practical guidelines in 3GPP TR 38.821 \cite{tgpp38821}. Only free-space path loss is considered, and small-scale fading is neglected. Normalizing the zenith high-gain received power to unity absorbs $P_s(\lambda_c/4\pi)^2G_{\mathrm{r}}$, so the carrier wavelength $\lambda_c$ does not enter the study.

\textit{Receiver positions and numerics.} The distribution is evaluated at zero longitude and the three CCDF latitudes of Table~\ref{tab:sim-params}. The moments are evaluated over the latitude sweep of the same table. The time-stepped simulation uses the two-body model to propagate the constellation in three dimensions along closed-form circular orbits, with the Earth constants, duration, and time step in Table~\ref{tab:sim-params}. All reported discrepancies use the same data as the figures.

\textit{Baselines.} BPP takes $K\sim\mathrm{Binomial}(N,\bar{p})$ visible satellites, where $\bar{p}=(1-\cos\gamma_{\max})/2$ is the area ratio of the user's visibility cap to the orbit sphere, and CPP follows the Poisson--Cox construction \cite{SG_CB_CPP} with Poisson$(N_o)$ planes, Poisson$(N_s)$ satellites per plane, and uniformly oriented plane normals. The NHPP \cite{SG_OtakiNHPP} instead assigns the satellites a latitude-dependent intensity while keeping their positions conditionally independent. 
The intensity is chosen as the first-order density of the Walker constellation, i.e., the image of the uniform measure on the phase torus, so it already encodes the altitude, the inclination, and the constellation size.

\subsubsection{Steady-State Distribution}
\label{sec:num-ss-dist}
% TODO: 对应中文稿 IV-D-2「稳态分布」

\begin{figure*}[t]
    \centering
    \includegraphics[width=\textwidth]{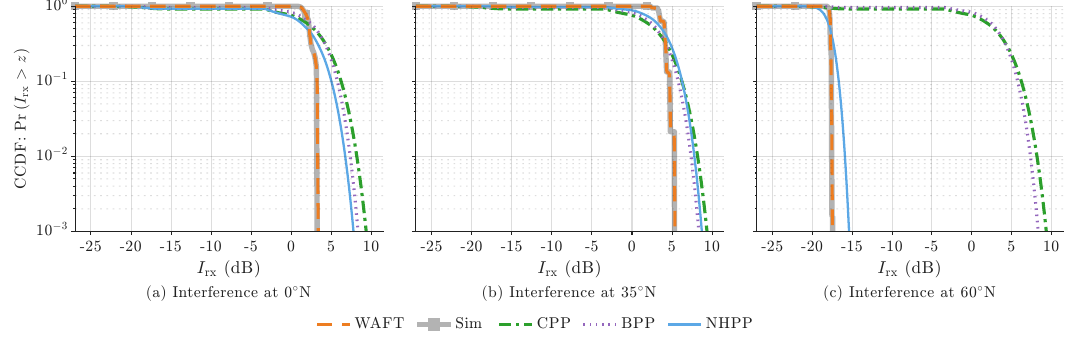}
    \caption{Steady-state CCDFs of the aggregate downlink interference for the Starlink Gen1 sub-constellation at $\theta_{\min}=10^\circ$.}
    \label{fig:ccdf}
\end{figure*}

% 提前声明矩图，使其紧随 CCDF 图进入双栏顶部浮动队列。
\begin{figure*}[!t]
    \centering
    \includegraphics[width=\textwidth]{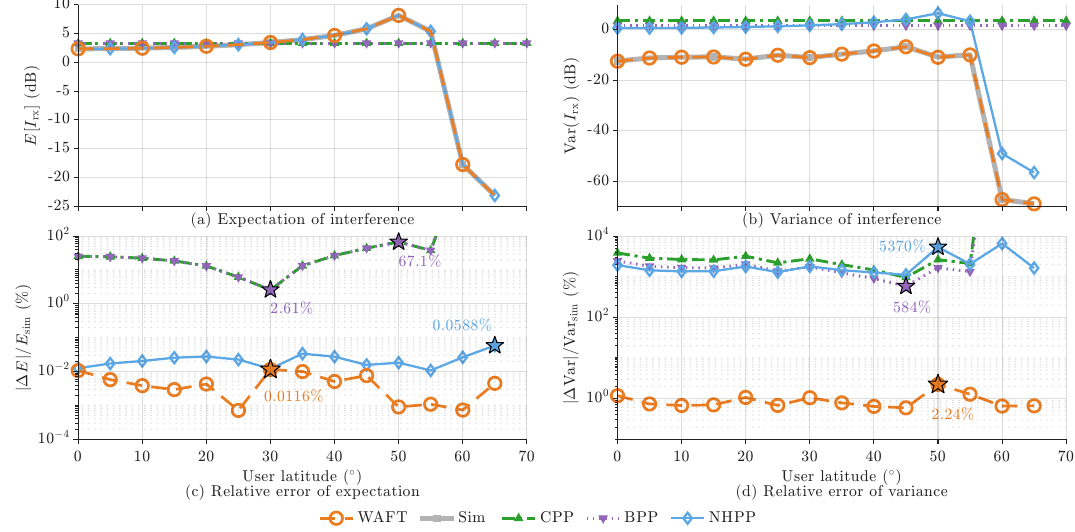}
    \caption{Steady-state moments of the aggregate downlink interference versus receiver latitude at $\theta_{\min}=10^\circ$: (a) expectation and (b) variance over $0^\circ$--$70^\circ$N, both in dB, where the variance is plotted as $10\log_{10}$ of $\mathrm{Var}(I_{\mathrm{rx}})$; (c)--(d) the corresponding relative discrepancies with respect to the simulation, in percent, whose vertical range is set by the data below $55^\circ$N.}
    \label{fig:moments}
\end{figure*}

Fig.~\ref{fig:ccdf} shows that the analytical CCDFs of the proposed method closely follow the time-stepped simulation baseline in all three panels, whereas the CCDFs of the isotropic BPP and CPP models are essentially independent of latitude and deviate substantially from the time-stepped simulation. 
The NHPP can reproduce the latitude dependence of the CCDF, but a significant discrepancy remains.

We quantify this discrepancy by the \textit{spurious probability mass}, i.e., the probability that an analytical model assigns to values outside the support bounds of the observable. Let $z_{\min}$ and $z_{\max}$ be the minimum and maximum of $I_{\mathrm{rx}}$ over the support set of the reference phase, estimated in practice from the extremes observed in the time-stepped simulation, and let $\bar F_{\mathrm{ana}}(\cdot)$ be the CCDF predicted by a given analytical model. The spurious probability mass is
\begin{equation}\label{eq:spm}
    \eta_{\mathrm{sp}} = \bigl[1-\bar F_{\mathrm{ana}}(z_{\min})\bigr] + \bar F_{\mathrm{ana}}(z_{\max}),
\end{equation}
i.e., the mass that $\bar F_{\mathrm{ana}}$ places below $z_{\min}$ or above $z_{\max}$. Because a finite time-stepped simulation samples only part of the trajectory, the estimated support bounds are inner approximations, so~\eqref{eq:spm} is a conservative upper bound.

Table~\ref{tab:spm} lists the resulting spurious probability masses of the four analytical models relative to the time-stepped simulation. Across the three latitudes, the spurious mass of WAFT is at most 1.11\%, whereas the mass of the three SG-based models ranges from 51.21\% to 99.00\%.

\begin{table}[htbp]
    \centering
    \caption{Spurious Probability Mass}
    \label{tab:spm}
    \begin{tabular}{ccccc}
        \hline
        Latitude & WAFT (\%) & CPP (\%) & BPP (\%) & NHPP (\%) \\
        \hline
        $0^\circ$  & 1.11 & 73.51 & 68.54 & 65.83 \\
        $35^\circ$ & 0.87 & 61.87 & 52.83 & 51.21 \\
        $60^\circ$ & 0.14 & 98.71 & 99.00 & 71.62 \\
        \hline
    \end{tabular}
\end{table}

\begin{remark}\label{rem:gibbs}
    The residual spurious mass of WAFT is a numerical artifact of the frequency truncation. Because $I_{\mathrm{rx}}(\boldsymbol{x})$ jumps both at the visibility boundary and at the $45^\circ$ gain threshold, its truncated spectral representation exhibits Gibbs oscillations that overshoot and undershoot the true support bounds, so the numerical CCDF assigns a small mass outside $[z_{\min},z_{\max}]$.
    A higher cutoff narrows these oscillations and hence reduces this mass, but does not eliminate it.
    At $60^\circ$N no satellite is visible in the high-gain regime, so only the weaker low-gain-to-invisible jump remains, which explains the smaller spurious mass at  $60^\circ$N.
\end{remark}

From an engineering perspective, the dynamic range of the interference is of primary interest. We therefore take the $0.1\%$ and $99.9\%$ percentiles of the steady-state distribution as the lower and upper percentile limits, respectively, and evaluate the discrepancies of both limits for the four analytical models relative to the finite-duration time-stepped simulation. Discrepancies are expressed in dB as the analytical value minus the simulated value, and are reported in Table~\ref{tab:bounds}.

\begin{table}[htbp]
    \centering
    \caption{Discrepancies of the Percentile Limits (dB)}
    \label{tab:bounds}
    \begin{tabular}{cccccc}
        \hline
        Latitude & Percentile & WAFT & CPP & BPP & NHPP \\
        \hline
        $0^\circ$  & $0.1\%$  & $-0.085$ & $-24.263$ & $-20.687$ & $-22.425$ \\
                   & $99.9\%$ & $+0.082$ & $+6.066$  & $+5.046$  & $+4.480$  \\[4pt]
        $35^\circ$ & $0.1\%$  & $-0.043$ & $-25.756$ & $-22.179$ & $-20.766$ \\
                   & $99.9\%$ & $+0.047$ & $+4.053$  & $+3.032$  & $+3.408$  \\[4pt]
        $60^\circ$ & $0.1\%$  & $-0.001$ & $-5.378$  & $-1.805$  & $-3.332$  \\
                   & $99.9\%$ & $-0.015$ & $+26.812$ & $+25.797$ & $+2.061$  \\
        \hline
    \end{tabular}
\end{table}

The absolute discrepancies of WAFT relative to the simulation remain within 0.09~dB for both limits. In contrast, the SG-based models place the lower percentile limit 1.8--25.8~dB below the simulated value and the upper percentile limit 2.0--26.8~dB above the simulated value, rendering the best-case interference overly optimistic and the worst-case interference overly conservative. Discrepancies of this magnitude can mislead spectrum-coexistence assessment, interference-coordination analysis, and link-margin design. Summing the two limits at each latitude, the isotropic BPP and CPP models exceed the simulated dynamic range by 25.2--32.2~dB. The NHPP narrows this excess to 5.4--26.9~dB, a reduction which comes from the intensity modelling the sparsity of the satellites at high latitudes. The dynamic-range discrepancy of WAFT remains 0.014--0.17~dB.

\subsubsection{Latitude Sweep of Moments}
\label{sec:num-moments}
% TODO: 对应中文稿 IV-D-3「矩的纬度扫描」
%   - 期望与方差随用户纬度的变化（图），及与 BPP/CPP 的相对误差对比。
\begin{figure}[!b]
    \centering
    \includegraphics{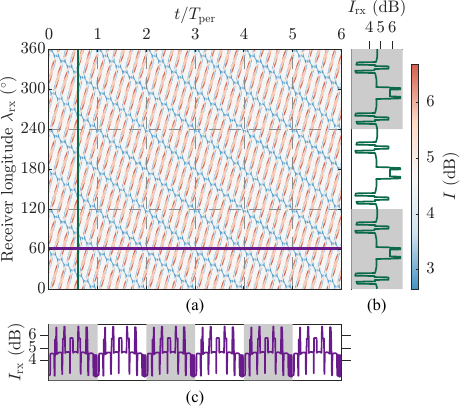}
    \caption{Instantaneous interference of the RGT constellation: (a) space--time interference over one sidereal day ($x$: $t/T_{\mathrm{per}}$, $y$: receiver longitude, color: $I_{\mathrm{rx}}$ in dB); (b) longitude profile at $t_0=0.6T_{\mathrm{per}}$; (c) time profile at $\lambda_0=61^\circ$. Gray bands mark one profile period; dashed lines mark $\Delta\lambda_{\mathrm{inst}}=120^\circ$ and $T_{\mathrm{per}}=3.989$~h.}
    \label{fig:recurr_validation}
\end{figure}

The two rows of Fig.~\ref{fig:moments} isolate the two ingredients that a model must reproduce.
The BPP and the CPP encode neither the anisotropy of the constellation nor any correlation among the satellites, so their moments are latitude-invariant with significant discrepancies.
The NHPP restores the first-order density, so the expectation agrees closely with the simulation, but the satellites remain conditionally independent and the variance is overestimated relative to the simulation.
The WAFT captures both ingredients.
The expectation agrees closely, and the residual discrepancy of its variance stems from the energy loss of the truncated spectrum and shrinks with a higher cutoff.
Above $60^\circ$N all visible satellites fall in the low-gain regime, and the interference is so weak that the CPP and the BPP are not included in the relative-discrepancy comparison.

\subsubsection{Validation for an RGT Constellation}
\label{sec:RGT-validation}

\begin{figure*}[!t]
    \centering
    \includegraphics[width=\textwidth]{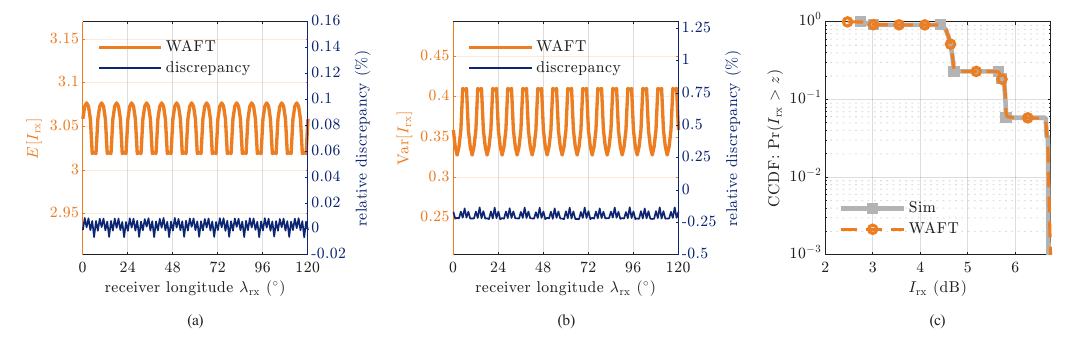}
    \caption{Steady-state validation for the RGT constellation: (a) expectation and (b) variance versus longitude, with signed discrepancies $100(X_{\mathrm{WAFT}}/X_{\mathrm{Sim}}-1)\%$; (c) CCDF at $\lambda_{\mathrm{rx}}=60^\circ$. Both moments exhibit an $8^\circ$ period.}
    \label{fig:recurr_steady}
\end{figure*}

The SG-based baselines compared in Section~\ref{sec:intf-validation} are omitted here, since such models describe the constellation by spatial snapshots and do not explicitly represent its temporal evolution, which is the focus of this subsection.

To verify the theory developed for RGT constellations, Table~\ref{tab:sim-params} defines a representative RGT Walker constellation, for which Remark~\ref{rem:temporal_period} predicts the temporal period $T_{\mathrm{per}}=3.989$~h, Theorem~\ref{theo:inst_period} the instantaneous longitude period $\Delta\lambda_{\mathrm{inst}}=120^\circ$ at any fixed time, and Theorem~\ref{theo:ss_period} the steady-state longitude period $\Delta\lambda_{\mathrm{steady}}=8^\circ$.

Fig.~\ref{fig:recurr_validation} shows the space--time map and representative profiles, verifying the temporal and instantaneous longitude periods. Fig.~\ref{fig:recurr_steady} compares the WAFT moments and CCDF with simulation, showing close agreement and the steady-state longitude period.

%% file: sections/conclusion.tex
\section{Conclusion}
\label{sec:conclusion}

This paper developed WAFT, a Fourier-analytic framework for steady-state analysis of additive observables in general Walker-$\delta$ constellations. The spectral representation separates constellation structure, observable form, and temporal evolution, yielding analytical expectation, variance, and CCDF expressions and closed-form instantaneous and steady-state longitude periods. WAFT accommodates arbitrary orbital altitude, inclination, constellation size, and phasing factor, covers both non-RGT and RGT constellations, and requires no satellite-position independence assumption.

For the non-RGT example (Starlink), WAFT exhibits percentile discrepancies below 0.09~dB relative to a finite-duration time-stepped simulation. For the RGT example, the temporal, instantaneous-longitude, and steady-state-longitude periods are verified.
Comparisons with the BPP, CPP, and NHPP examine the effects of intensity matching and of the conditional independence of satellite positions on the SG discrepancy.

The WAFT framework assumes ideal circular orbits, a spherical Earth, and a single-layer constellation. 
Future work includes multi-shell and heterogeneous constellations, temporal correlation, and EPFD-compliance workflows.

%% file: sections/appendix.tex
%% ============================================================
%%  附录（Appendices）  ——  sections/appendix.tex
%%  关联方式：main.tex 中  \appendices  +  \input{sections/appendix}
%%  说明：
%%   - \appendices 之后，每个 \section 会自动编号为 APPENDIX A、B、…
%%   - 附录内的公式默认「延续正文编号」（正文末式若为 (1)，附录首式即 (2)）；
%%     这是 IEEE 指南首选做法。若坚持 (A1)、(A2) 形式，可在此文件顶部加入：
%%       \renewcommand{\theequation}{A\arabic{equation*}}\setcounter{equation*}{0}
%%   - 增删附录 = 增删 \section；把下面的 TODO 替换为真实证明即可。
%% ============================================================

\section{Proof of Lemma~\ref{lem:visibility}}
\label{app:lemma1}
\begin{IEEEproof}
By~\eqref{eq:theta_of_gamma} and the monotonicity of $\theta$ in $\gamma$ established in Section~\ref{sec:visibility}, $\theta(\Omega,u)\ge\theta_{\min}$ is equivalent to $\cos\gamma(\Omega,u)\ge\cos\gamma_{\max}$.
Expanding~\eqref{eq:gamma_def} with the unit satellite position vector~\eqref{coordinates} yields
\begin{equation*}
    \begin{aligned}
        \cos\gamma(\Omega,u) &= A(\Omega)\cos u + B(\Omega)\sin u\\
        &= R(\Omega)\cos\bigl(u-\phi_0(\Omega)\bigr),
    \end{aligned}
\end{equation*}
where $A(\Omega)=\cos\varphi_{\mathrm{rx}}\cos(\Omega-\lambda_{\mathrm{rx}})$, $B(\Omega)=\sin\varphi_{\mathrm{rx}}\sin i-\cos\varphi_{\mathrm{rx}}\sin(\Omega-\lambda_{\mathrm{rx}})\cos i$, $R(\Omega)=\sqrt{A^2(\Omega)+B^2(\Omega)}$, and $\phi_0(\Omega)=\operatorname{atan2}(B(\Omega),A(\Omega))$, defined when $R(\Omega)>0$.

Expanding $R^2=A^2+B^2$ and substituting $\cos\beta(\Omega)=\hat{\boldsymbol{r}}_{\mathrm{rx}}^{\mathsf{T}}\boldsymbol{n}(\Omega)$ yields $R^2(\Omega)=1-\cos^2\beta(\Omega)$, hence $R(\Omega)=\sin\beta(\Omega)$.
Since the left-hand side of $R(\Omega)\cos(u-\phi_0(\Omega))\ge\cos\gamma_{\max}$ attains its maximum $R(\Omega)$ at $u=\phi_0(\Omega)$, a solution exists if and only if
\begin{equation}\label{eq:app_beta_interval}
    \sin\beta(\Omega)\ge\cos\gamma_{\max}
    \iff \beta(\Omega)\in\bigl[\tfrac{\pi}{2}-\gamma_{\max},\ \tfrac{\pi}{2}+\gamma_{\max}\bigr].
\end{equation}

Under~\eqref{eq:app_beta_interval}, $\sin\beta(\Omega)>0$ and $\cos\gamma_{\max}/\sin\beta(\Omega)\in(0,1]$, so the condition reduces to
\begin{equation*}
    \cos\bigl(u-\phi_0(\Omega)\bigr)\ge\frac{\cos\gamma_{\max}}{\sin\beta(\Omega)}=\cos\Psi(\Omega),
\end{equation*}
i.e., $u\in[\phi_0(\Omega)-\Psi(\Omega),\ \phi_0(\Omega)+\Psi(\Omega)]$ $(\mathrm{mod}\ 2\pi)$, with $\Psi$ as in~\eqref{eq:interval_halfwidth}.
When $\sin\beta(\Omega)<\cos\gamma_{\max}$, \eqref{eq:app_beta_interval} has no solution and the visible set is empty. Accordingly, $\Psi(\Omega)$ is set to $0$ by default in this case.
\end{IEEEproof}

\section{Proof of Theorem~\ref{theo:spectral_rep}}
\label{app:theo1}
\begin{IEEEproof}
    Represent the constellation by the measure $\Delta_{\boldsymbol{x}}=\sum_{i_p,j_s}\delta(\Omega-\Omega_{i_p},u-u_{i_pj_s})$ on $\mathbb{T}^2$, whose atoms are the $N$ points $\boldsymbol{s}_n=\boldsymbol{x}+\boldsymbol{\lambda}_n\pmod{2\pi}$, $n=(i_p,j_s)$, with $\boldsymbol{\lambda}_n=i_p\boldsymbol{x}_1+j_s\boldsymbol{x}_2$ and $\boldsymbol{x}_1,\boldsymbol{x}_2$ the generators of $\mathcal{L}$ in \eqref{eq:phase_lattice}. Pairing this measure with the observation field gives $Z_{\mathrm{rx}}=\iint_{\mathbb{T}^2}U_{\mathrm{rx}}\,\mathrm{d}\Delta_{\boldsymbol{x}}$. With the torus Fourier transform $\hat{g}(\boldsymbol{y})=\iint_{\mathbb{T}^2}g(\boldsymbol{\xi})\,e^{-j\boldsymbol{y}^{\mathsf{T}}\boldsymbol{\xi}}\,\mathrm{d}\boldsymbol{\xi}$ and inversion $g(\boldsymbol{\xi})=\frac{1}{4\pi^2}\sum_{\boldsymbol{y}\in\mathbb{Z}^2}\hat{g}(\boldsymbol{y})e^{j\boldsymbol{y}^{\mathsf{T}}\boldsymbol{\xi}}$, the Fourier transform of the measure $\Delta_{\boldsymbol{x}}$ is
    \begin{equation*}
    \hat{\Delta}_{\boldsymbol{x}}(\boldsymbol{y})=e^{-j\boldsymbol{y}^{\mathsf{T}}\boldsymbol{x}}\sum_{i_p,j_s}e^{-j\boldsymbol{y}^{\mathsf{T}}\boldsymbol{\lambda}_n}
    =\begin{cases}N e^{-j\boldsymbol{y}^{\mathsf{T}}\boldsymbol{x}}, & \boldsymbol{y}\in\mathcal{L}^\perp,\\ 0, & \text{otherwise},\end{cases}
    \end{equation*}
    Since $Z_{\mathrm{rx}}=\iint U_{\mathrm{rx}}\,\mathrm{d}\Delta_{\boldsymbol{x}}$ is the zero-th Fourier coefficient of the product measure $U_{\mathrm{rx}}\Delta_{\boldsymbol{x}}$, the convolution property on torus $\widehat{(fg)}(\boldsymbol{y})=\frac{1}{4\pi^2}\sum_{\boldsymbol{y}'\in\mathbb{Z}^2}\hat f(\boldsymbol{y}')\hat g(\boldsymbol{y}-\boldsymbol{y}')$ gives
    \begin{equation*}
    Z_{\mathrm{rx}}=\frac{1}{4\pi^2}\sum_{\boldsymbol{y}\in\mathbb{Z}^2}\hat U_{\mathrm{rx}}(\boldsymbol{y})\,\hat\Delta_{\boldsymbol{x}}(-\boldsymbol{y})
    =\frac{N}{4\pi^2}\sum_{\boldsymbol{y}\in\mathcal{L}^\perp}\hat U_{\mathrm{rx}}(\boldsymbol{y})e^{j\boldsymbol{y}^{\mathsf{T}}\boldsymbol{x}},
    \end{equation*}
    where $\hat\Delta_{\boldsymbol{x}}(-\boldsymbol{y})=N e^{j\boldsymbol{y}^{\mathsf{T}}\boldsymbol{x}}$ on $\mathcal{L}^\perp$ and zero elsewhere, by the symmetry of the dual lattice. This is \eqref{eq:spectral_rep}. Since $U_{\mathrm{rx}}$ is bounded on $\mathbb{T}^2$, it belongs to $L^2(\mathbb{T}^2)$, and its Fourier coefficients are square-summable by Parseval's identity. Restricting the coefficients to $\mathcal{L}^\perp$ preserves square-summability, so the series converges in $L^2(\mathbb{T}^2)$.
\end{IEEEproof}

\section{Proof of Theorem~\ref{theo:spectral_rep_recurr}}
\label{app:theo2}
\begin{IEEEproof}
    Substitute the trajectory $\boldsymbol{x}(t)=\boldsymbol{x}(0)+\boldsymbol{v}t$, with $\boldsymbol{v}=(-\omega_E,\,(p/q)\omega_E)$, into \eqref{eq:spectral_rep_with_t}. Writing $\boldsymbol{y}=(a,b)$ and $l\triangleq pb-qa$,
    \begin{equation*}
    \boldsymbol{y}^{\mathsf{T}}\boldsymbol{x}(t)
    =\boldsymbol{y}^{\mathsf{T}}\boldsymbol{x}(0)+\frac{pb-qa}{q}\omega_E t
    =\boldsymbol{y}^{\mathsf{T}}\boldsymbol{x}(0)+\frac{l\,\omega_E}{q}t,
    \end{equation*}
    so all terms with the same $l$ share the temporal frequency $l\omega_E/q$. Collecting them gives \eqref{eq:spectral_rep_recur_t} with the coefficients \eqref{eq:Hl}.

    Note that $\boldsymbol{y}=(mN_o-Fn,\,nN_s)$ on $\mathcal{L}^\perp$ yields $l=n(pN_s+qF)-qmN_o$, so $l$ ranges exactly over $g_t\mathbb{Z}$, with $g_t=\gcd(pN_s+qF,\,qN_o)$. Under the coefficient bound in part (b), absolute convergence within each frequency group justifies the collection. Indeed, for fixed $l$, the points satisfying $pb-qa=l$ form a one-dimensional affine sublattice; its index $k$ satisfies $\lvert\boldsymbol{y}\rvert\ge c(1+\lvert k\rvert)$ for some $c>0$, and hence
    \begin{equation*}
    \sum_{pb-qa=l}\lvert\hat U_0(\boldsymbol{y})\rvert\le C\sum_{k\in\mathbb{Z}}(1+\lvert k\rvert)^{-3/2}<\infty.
    \end{equation*}
    Thus every $\hat H_l$ is well defined. Moreover, the time trace is bounded because it is a sum of $N$ bounded field values, and therefore belongs to $L^2$ over one trajectory period. Its Fourier coefficients are precisely $\zeta\hat H_l$ by the grouping above; orthogonality of the time exponentials then gives convergence of \eqref{eq:spectral_rep_recur_t} to $Z_{\mathrm{rx}}(\boldsymbol{x}(t))$ in $L^2$ over one trajectory period.
\end{IEEEproof}

\section{Proof of Theorem~\ref{theo:inst_period}}
\label{app:theo3}
\begin{IEEEproof}
    By Theorem~\ref{theo:spectral_rep} and the longitude translation $U_{\mathrm{rx}}(\Omega,u)=U_0(\Omega-\lambda_{\mathrm{rx}},u)$,
    \begin{equation*}
        Z_{\mathrm{rx}}(\lambda_{\mathrm{rx}})=\zeta\sum_{(a,b)\in\mathcal{L}^\perp}\hat U_0(a,b)\,e^{j(a\Omega_x+b u_x)}\,e^{-ja\lambda_{\mathrm{rx}}},
    \end{equation*}
    with convergence in $L^2$. Only the harmonics $e^{-ja\lambda_{\mathrm{rx}}}$ depend on longitude, and $a=mN_o-Fn\in\gcd(N_o,F)\mathbb{Z}$.

    Fourier uniqueness implies that a shift $\delta$ is a period if and only if $a\delta\in2\pi\mathbb{Z}$ for every $a$ having a nonzero dual-lattice coefficient. With $G_\lambda$ as defined in Theorem~\ref{theo:inst_period}, the period set is therefore $(2\pi/G_\lambda)\mathbb{Z}$, giving the minimum positive period $2\pi/G_\lambda$ under the stated nondegeneracy assumption. Every such $a$ is divisible by $\gcd(N_o,F)$. If a nonzero dual-lattice coefficient with $a=\pm\gcd(N_o,F)$ exists, divisibility in both directions gives $G_\lambda=\gcd(N_o,F)$.
\end{IEEEproof}

\section{Proof of Theorem~\ref{theo:ss_period}}
\label{app:theo4}
\begin{IEEEproof}
    In an RGT constellation, each satellite moves along a closed geodesic $c=p\Omega+qu$ of \eqref{eq:geodesic}, so the steady-state observable depends only on the relative position between the set of trajectory lines and the receiver observation field.
    A receiver-longitude shift $\delta$ is equivalent to shifting the LAN of every satellite by $-\delta$, i.e., $c\mapsto c-p\delta$. 

    For the $j_s$-th satellite in the $i_p$-th plane,
    \begin{equation*}
        c_{i_pj_s}=c_0+\frac{2\pi}{N}\bigl[i_p(pN_s+qF)+j_sqN_o\bigr]\pmod{2\pi},
    \end{equation*}
    with $c_0=p\Omega_x(0)+qu_x(0)$, so the trajectory parameters form the uniform lattice generated by $2\pi g_c/N$, where $g_c=\gcd(pN_s+qF,\,qN_o,\,N)$. The minimum translation of this trajectory set along $c$ is $\Delta c_{\min}=2\pi g_c/N$, and $p\delta=\Delta c_{\min}$ gives
    \begin{equation*}
        \delta=\frac{2\pi g_c}{pN}
        =\frac{2\pi}{pN}\gcd\bigl(pN_s+qF,\ qN_o,\ N\bigr).
    \end{equation*}
    Hence this shift is always a longitude period of the steady-state distribution. Under the stated assumption that distinct trajectory-set translates induce distinct distributions, minimality of the trajectory-set translation also implies minimality of the distribution period.
\end{IEEEproof}

\section{Proof of Corollary~\ref{cor:period_ratio}}
\label{app:cor1}
\begin{IEEEproof}
    Let $d=\gcd(N_o,F)$ and $\bar N_o=N_o/d$. By the definition of $g_c$, we have $g_c\mid pN_s+qF$ and $g_c\mid qN_o$. Multiplying the first divisibility by $\bar N_o$ and using the second gives $g_c\mid pN_s\bar N_o$. Therefore,
    \begin{equation*}
        \frac{\Delta\lambda_{\mathrm{inst}}}{\Delta\lambda_{\mathrm{steady}}}
        = \frac{2\pi/d}{2\pi g_c/(pN)}
        = \frac{pN}{d g_c}
        = \frac{pN_s\bar N_o}{g_c}\in\mathbb{Z}^{+},
    \end{equation*}
    where the nondegeneracy assumption supplies $\Delta\lambda_{\mathrm{inst}}=2\pi/d$.
\end{IEEEproof}

\section{Proof of Theorem~\ref{theo:moments_for_non_recurr}}
\label{app:theo5}
\begin{IEEEproof}
    For the non-RGT case, the steady-state average is over $\boldsymbol{x}$ uniform on $\mathbb{T}^2$. By the orthogonality of the torus exponentials, for any $\boldsymbol{y}\in\mathbb{Z}^2$,
    \begin{equation}\label{eq:torus_orthogonality}
        \mathbb{E}\bigl[e^{j\boldsymbol{y}^{\mathsf{T}}\boldsymbol{x}}\bigr]
        = \frac{1}{4\pi^2}\iint_{\mathbb{T}^2} e^{j\boldsymbol{y}^{\mathsf{T}}\boldsymbol{x}}\,\mathrm{d}\boldsymbol{x}
        = \mathds{1}\{\boldsymbol{y}=\boldsymbol{0}\}.
    \end{equation}
    Since \eqref{eq:spectral_rep} converges in $L^2(\mathbb{T}^2)$ and the expectation is a continuous functional there, the sum and the expectation may be interchanged:
    \begin{equation*}
        \mathbb{E}[Z_{\mathrm{rx}}]=\zeta\sum_{\boldsymbol{y}\in\mathcal{L}^\perp}\hat U_{\mathrm{rx}}(\boldsymbol{y})\,\mathbb{E}\bigl[e^{j\boldsymbol{y}^{\mathsf{T}}\boldsymbol{x}}\bigr]=\zeta\,\hat U_{\mathrm{rx}}(\boldsymbol{0}).
    \end{equation*}
    For the second central moment, expand $\lvert Z_{\mathrm{rx}}-\mathbb{E}Z_{\mathrm{rx}}\rvert^2=(Z_{\mathrm{rx}}-\mathbb{E}Z_{\mathrm{rx}})\overline{(Z_{\mathrm{rx}}-\mathbb{E}Z_{\mathrm{rx}})}$. The $L^2$ convergence permits termwise expansion, and \eqref{eq:torus_orthogonality} collapses the double sum onto $\boldsymbol{y}_1=\boldsymbol{y}_2$:
    \begin{equation*}
        \begin{aligned}
            \mathrm{Var}(Z_{\mathrm{rx}})
            & =\zeta^2\sum_{\boldsymbol{y}\in\mathcal{L}^\perp}\bigl|\hat U_{\mathrm{rx}}(\boldsymbol{y})\bigr|^2-\zeta^2\bigl|\hat U_{\mathrm{rx}}(\boldsymbol{0})\bigr|^2\\
            & =\zeta^2\sum_{\boldsymbol{y}\in\mathcal{L}^\perp\setminus\{\boldsymbol{0}\}}\bigl|\hat U_{\mathrm{rx}}(\boldsymbol{y})\bigr|^2.
        \end{aligned}
    \end{equation*}
\end{IEEEproof}

%% file: refs.bib
@article{Arch_KeyTech,
  author   = {Su, Yongtao and Liu, Yaoqi and Zhou, Yiqing and Yuan, Jinhong and Cao, Huan and Shi, Jinglin},
  journal  = {IEEE Wireless Communications},
  title    = {Broadband {LEO} Satellite Communications: Architectures and Key Technologies},
  year     = {2019},
  volume   = {26},
  number   = {2},
  pages    = {55-61},
  doi      = {10.1109/MWC.2019.1800299}
}

@article{CB_WPP_1,
  author   = {Choi, Chang-Sik and Baccelli, Francois},
  journal  = {IEEE Transactions on Vehicular Technology},
  title    = {Stochastic Geometry and Dynamical System Analysis of {Walker} Satellite Constellations},
  year     = {2026},
  volume   = {75},
  number   = {3},
  pages    = {5127-5132},
  doi      = {10.1109/TVT.2025.3608470}
}

@misc{CB_WPP_2,
  author       = {Choi, Chang-Sik and Baccelli, Francois},
  title        = {Dynamical System Characterization of Heterogeneous {Walker} Satellite Networks: An Orbit-Aware Stochastic Geometry Perspective},
  howpublished = {arXiv:2606.29433},
  month        = jun,
  year         = {2026},
  doi          = {10.48550/arXiv.2606.29433}
}

@article{CmparOf3LEO,
  title   = {A technical comparison of three low {Earth} orbit satellite constellation systems to provide global broadband},
  journal = {Acta Astronautica},
  volume  = {159},
  pages   = {123-135},
  year    = {2019},
  issn    = {0094-5765},
  doi     = {https://doi.org/10.1016/j.actaastro.2019.03.040},
  author  = {Inigo {del Portillo} and Bruce G. Cameron and Edward F. Crawley}
}

@inproceedings{DelyNotOption,
  author    = {Handley, Mark},
  title     = {Delay is Not an Option: Low Latency Routing in Space},
  year      = {2018},
  isbn      = {9781450361200},
  publisher = {Association for Computing Machinery},
  address   = {New York, NY, USA},
  doi       = {10.1145/3286062.3286075},
  booktitle = {Proceedings of the 17th ACM Workshop on Hot Topics in Networks},
  pages     = {85-91},
  numpages  = {7},
  location  = {Redmond, WA, USA},
  series    = {HotNets '18}
}

@book{dynamical_theory,
  place      = {Cambridge},
  title      = {Introduction to the Modern Theory of Dynamical Systems},
  publisher  = {Cambridge University Press},
  author     = {Katok, Anatole and Hasselblatt, Boris},
  year       = {1995}
}

@article{FiveGFrmSpce,
  author   = {Lin, Xingqin and Rommer, Stefan and Euler, Sebastian and Yavuz, Emre A. and Karlsson, Robert S.},
  journal  = {IEEE Communications Standards Magazine},
  title    = {{5G} from Space: An Overview of {3GPP} Non-Terrestrial Networks},
  year     = {2021},
  volume   = {5},
  number   = {4},
  pages    = {147-153},
  doi      = {10.1109/MCOMSTD.011.2100038}
}

@manual{itu_rr2024_art22,
  organization = {International Telecommunication Union},
  title        = {Radio Regulations, Volume 1: Articles, Edition of 2024},
  address      = {Geneva, Switzerland},
  year         = {2024},
  note         = {{A}rticle 22, Section II, No. 22.5C and Table 22-1A;
                  in force as from 1 January 2025}
}

@misc{itu2024_wrc23,
  author       = {{International Telecommunication Union}},
  title        = {Final Acts of the {World Radiocommunication Conference} ({WRC-23})},
  howpublished = {International Telecommunication Union},
  address      = {Geneva, Switzerland},
  year         = {2024}
}

@misc{itur2023s1503,
  author       = {{ITU-R}},
  title        = {Functional description to be used in developing software tools for determining conformity of non-geostationary-satellite orbit fixed-satellite service systems or networks with limits contained in {Article 22} of the {Radio Regulations}},
  howpublished = {Recommendation ITU-R S.1503-4},
  month        = sep,
  year         = {2023}
}

@inproceedings{NewCmparOf4LEO,
  author    = {Pachler, Nils and del Portillo, Inigo and Crawley, Edward F. and Cameron, Bruce G.},
  booktitle = {2021 IEEE International Conference on Communications Workshops (ICC Workshops)},
  title     = {An Updated Comparison of Four Low {Earth} Orbit Satellite Constellation Systems to Provide Global Broadband},
  year      = {2021},
  volume    = {},
  number    = {},
  pages     = {1-7},
  doi       = {10.1109/ICCWorkshops50388.2021.9473799}
}

@article{NewSpaceEra,
  author   = {Kodheli, Oltjon and Lagunas, Eva and Maturo, Nicola and Sharma, Shree Krishna and Shankar, Bhavani and Montoya, Jesus Fabian Mendoza and Duncan, Juan Carlos Merlano and Spano, Danilo and Chatzinotas, Symeon and Kisseleff, Steven and Querol, Jorge and Lei, Lei and Vu, Thang X. and Goussetis, George},
  journal  = {IEEE Communications Surveys \& Tutorials},
  title    = {Satellite Communications in the New Space Era: A Survey and Future Challenges},
  year     = {2021},
  volume   = {23},
  number   = {1},
  pages    = {70-109},
  doi      = {10.1109/COMST.2020.3028247}
}

@article{Quick2Dmap,
  title   = {Quick coverage analysis of mega {Walker} Constellation based on {2D} map},
  journal = {Acta Astronautica},
  volume  = {188},
  pages   = {99-109},
  year    = {2021},
  issn    = {0094-5765},
  doi     = {https://doi.org/10.1016/j.actaastro.2021.07.008},
  author  = {Yupeng Gong and Shijie Zhang and Xuan Peng}
}

@article{SG_4LEO,
  author   = {Talgat, Anna and Kishk, Mustafa A. and Alouini, Mohamed-Slim},
  journal  = {IEEE Communications Letters},
  title    = {Stochastic Geometry-Based Analysis of {LEO} Satellite Communication Systems},
  year     = {2021},
  volume   = {25},
  number   = {8},
  pages    = {2458-2462},
  doi      = {10.1109/LCOMM.2020.3029808}
}

@book{SG_4WN,
  place     = {Cambridge},
  title     = {Stochastic Geometry for Wireless Networks},
  publisher = {Cambridge University Press},
  author    = {Haenggi, Martin},
  year      = {2012},
  doi       = {10.1017/CBO9781139043816}
}

@article{SG_alhourani2021analytic,
  author   = {Al-Hourani, Akram},
  journal  = {IEEE Wireless Communications Letters},
  title    = {An Analytic Approach for Modeling the Coverage Performance of Dense Satellite Networks},
  year     = {2021},
  volume   = {10},
  number   = {4},
  pages    = {897-901},
  doi      = {10.1109/LWC.2021.3049179}
}

@article{SG_BPPNearest,
  author  = {Talgat, Anna and Kishk, Mustafa A. and Alouini, Mohamed-Slim},
  journal = {IEEE Communications Letters},
  title   = {Nearest Neighbor and Contact Distance Distribution for Binomial Point Process on Spherical Surfaces},
  year    = {2020},
  volume  = {24},
  number  = {12},
  pages   = {2659-2663},
  doi     = {10.1109/LCOMM.2020.3019436}
}

@article{SG_CB_CPP,
  author   = {Choi, Chang-Sik and Baccelli, François},
  journal  = {IEEE Transactions on Communications},
  title    = {A Novel Analytical Model for {LEO} and {MEO} Satellite Networks Based on {Cox} Point Processes},
  year     = {2025},
  volume   = {73},
  number   = {4},
  pages    = {2265-2279},
  doi      = {10.1109/TCOMM.2024.3471978}
}

@article{SG_CB_CPP4HetLEO,
  author   = {Choi, Chang-Sik and Baccelli, François},
  journal  = {IEEE Transactions on Vehicular Technology},
  title    = {{Cox} Point Processes for Multi Altitude {LEO} Satellite Networks},
  year     = {2024},
  volume   = {73},
  number   = {10},
  pages    = {15916-15921},
  doi      = {10.1109/TVT.2024.3414833}
}

@article{SG_Choi_HetLEO,
  author   = {Choi, Chang-Sik},
  journal  = {IEEE Transactions on Wireless Communications},
  title    = {Modeling and Analysis of Downlink Communications in a Heterogeneous {LEO} Satellite Network},
  year     = {2024},
  volume   = {23},
  number   = {8},
  pages    = {8588-8602},
  doi      = {10.1109/TWC.2024.3351876}
}

@article{SG_constraint_angle,
  author   = {He, Yuan and Li, Zhuoyang and Chen, Yiqi},
  journal  = {IEEE Internet of Things Journal},
  title    = {Tractable Modeling and Performance Analysis of Low-{Earth} Orbit Satellite Constellations},
  year     = {2024},
  volume   = {11},
  number   = {17},
  pages    = {28297-28306},
  doi      = {10.1109/JIOT.2024.3402140}
}

@article{SG_evaluate,
  author   = {Wang, Ruibo and Kishk, Mustafa A. and Alouini, Mohamed-Slim},
  journal  = {IEEE Communications Letters},
  title    = {Evaluating the Accuracy of Stochastic Geometry Based Models for {LEO} Satellite Networks Analysis},
  year     = {2022},
  volume   = {26},
  number   = {10},
  pages    = {2440-2444},
  doi      = {10.1109/LCOMM.2022.3194210}
}

@article{SG_OrbGeo,
  author   = {Lee, Junse and Noh, Song and Jung, Sooyeob and Lee, Namyoon},
  journal  = {IEEE Access},
  title    = {Coverage Analysis of {LEO} Satellite Downlink Networks: Orbital Geometry Dependent Approach},
  year     = {2024},
  volume   = {12},
  number   = {},
  pages    = {196939-196953},
  doi      = {10.1109/ACCESS.2024.3522377}
}

@inproceedings{SG_Otaki_model_NHPP,
  author    = {Okati, Niloofar and Riihonen, Taneli},
  booktitle = {2021 IEEE 93rd Vehicular Technology Conference (VTC2021-Spring)},
  title     = {Modeling and Analysis of {LEO} Mega-Constellations as Nonhomogeneous {Poisson} Point Processes},
  year      = {2021},
  volume    = {},
  number    = {},
  pages     = {1-5},
  doi       = {10.1109/VTC2021-Spring51267.2021.9448844}
}

@article{SG_OtakiNHPP,
  author   = {Okati, Niloofar and Riihonen, Taneli},
  journal  = {IEEE Transactions on Communications},
  title    = {Nonhomogeneous Stochastic Geometry Analysis of Massive {LEO} Communication Constellations},
  year     = {2022},
  volume   = {70},
  number   = {3},
  pages    = {1848-1860},
  doi      = {10.1109/TCOMM.2022.3143131}
}

@article{SG_OtakiSG,
  author   = {Okati, Niloofar and Riihonen, Taneli and Korpi, Dani and Angervuori, Ilari and Wichman, Risto},
  journal  = {IEEE Transactions on Communications},
  title    = {Downlink Coverage and Rate Analysis of Low {Earth} Orbit Satellite Constellations Using Stochastic Geometry},
  year     = {2020},
  volume   = {68},
  number   = {8},
  pages    = {5120-5134},
  doi      = {10.1109/TCOMM.2020.2990993}
}

@article{SG_PAHPP,
  author   = {Zhang, Haoxing and Miao, Xiaqing and Ni, Zihan and Wang, Shuai and Pan, Gaofeng and Cavdar, Cicek and An, Jianping},
  journal  = {IEEE Transactions on Wireless Communications},
  title    = {{LEO} Mega-Constellation-Terrestrial Communications Suffering {Poisson} Arc Hardcore Distributed Space Interference},
  year     = {2025},
  volume   = {24},
  number   = {4},
  pages    = {2707-2721},
  doi      = {10.1109/TWC.2024.3523950}
}

@article{SG_Shadow,
  author   = {Jung, Dong-Hyun and Ryu, Joon-Gyu and Byun, Woo-Jin and Choi, Junil},
  journal  = {IEEE Transactions on Communications},
  title    = {Performance Analysis of Satellite Communication System Under the Shadowed-{Rician} Fading: A Stochastic Geometry Approach},
  year     = {2022},
  volume   = {70},
  number   = {4},
  pages    = {2707-2721},
  doi      = {10.1109/TCOMM.2022.3142290}
}

@article{SG_Survey,
  author   = {Wang, Ruibo and Kishk, Mustafa A. and Alouini, Mohamed-Slim},
  journal  = {IEEE Communications Surveys \& Tutorials},
  title    = {Modeling and Analysis of Non-Terrestrial Networks by Spherical Stochastic Geometry: A Survey},
  year     = {2026},
  volume   = {28},
  pages    = {1879-1905},
  doi      = {10.1109/COMST.2025.3562489}
}

@article{SG_wang2022conditional,
  author   = {Wang, Ruibo and Talgat, Anna and Kishk, Mustafa A. and Alouini, Mohamed-Slim},
  journal  = {IEEE Communications Letters},
  title    = {Conditional Contact Angle Distribution in {LEO} Satellite-Relayed Transmission},
  year     = {2022},
  volume   = {26},
  number   = {11},
  pages    = {2735-2739},
  doi      = {10.1109/LCOMM.2022.3196152}
}

@techreport{walker1970_circular,
  author      = {J. G. Walker},
  title       = {Circular Orbit Patterns Providing Continuous Whole {Earth} Coverage},
  institution = {Royal Aircraft Establishment},
  type        = {Technical Report},
  number      = {70211},
  month       = nov,
  year        = {1970}
}

@article{weyl1916gleichverteilung,
  title     = {{\"U}ber die {G}leichverteilung von {Z}ahlen mod.\ {E}ins},
  author    = {Weyl, Hermann},
  journal   = {Mathematische Annalen},
  volume    = {77},
  number    = {3},
  pages     = {313--352},
  year      = {1916},
  publisher = {Springer}
}

@INPROCEEDINGS{Starlink_phasing_factor_17,
  author={Liang, Jintao and Chaudhry, Aizaz U. and Yanikomeroglu, Halim},
  booktitle={2021 IEEE 4th 5G World Forum (5GWF)}, 
  title={Phasing Parameter Analysis for Satellite Collision Avoidance in Starlink and Kuiper Constellations}, 
  year={2021},
  volume={},
  number={},
  pages={493-498},
  doi={10.1109/5GWF52925.2021.00093}}

@misc{tgpp38821,
  author       = {{3GPP}},
  title        = {Solutions for {NR} to Support Non-Terrestrial Networks ({NTN})},
  howpublished = {3GPP TR 38.821},
  month        = dec,
  year         = {2019}
}

@article{Kuang_AggInt_AAST,
  author  = {Lin, Ziqiao and Jin, Jin and Yan, Jian and Kuang, Linling},
  title   = {A Method for Calculating the Probability Distribution of Interference Involving Mega-Constellations},
  journal = {Advances in Astronautics Science and Technology},
  volume  = {4},
  number  = {1},
  pages   = {107--117},
  year    = {2021},
  doi     = {10.1007/s42423-021-00079-0}
}

@article{Kuang_UplinkInt_IoTJ,
  author  = {Jia, Haoge and Ni, Zuyao and Jiang, Chunxiao and Kuang, Linling and Lu, Jianhua},
  title   = {Uplink Interference and Performance Analysis for Megasatellite Constellation},
  journal = {IEEE Internet of Things Journal},
  volume  = {9},
  number  = {6},
  pages   = {4318--4329},
  year    = {2022},
  doi     = {10.1109/JIOT.2021.3104095}
}

@article{Kuang_BlindSpot_TAES,
  author  = {Zhang, Chi and Jin, Jin and Kuang, Linling and Jiang, Chunxiao and He, Yuanzhi},
  title   = {Blind Spot of Spectrum Awareness Techniques in Nongeostationary Satellite Systems},
  journal = {IEEE Transactions on Aerospace and Electronic Systems},
  volume  = {54},
  number  = {6},
  pages   = {3150--3159},
  year    = {2018},
  doi     = {10.1109/TAES.2018.2826258}
}
